\documentclass[a4paper,twocolumn,showpacs,superscriptaddress,floatfix,nofootinbib]{revtex4-1}

\usepackage{latexsym}
\usepackage{amssymb}
\usepackage{amsfonts}
\usepackage{amsmath}
\usepackage{bm}
\usepackage{graphicx}   
\usepackage{color}
\usepackage{subfigure}
\usepackage{times}
\usepackage{units}
\usepackage{hyperref}

\graphicspath{ {./plots/} }

\begin{document}

\title[Long-lived BNS merger remnants]{First 100\,ms of a long-lived magnetized neutron star\\formed in a binary neutron star merger}

\author{Riccardo Ciolfi}
\address{INAF, Osservatorio Astronomico di Padova, Vicolo dell'Osservatorio 5, I-35122 Padova, Italy}
\address{INFN, Sezione di Padova, Via Francesco Marzolo 8, I-35131 Padova, Italy}

\author{Wolfgang Kastaun}
\address{Max Planck Institute for Gravitational Physics (Albert
  Einstein Institute), Callinstrasse 38, 30167 Hannover, Germany}
\address{Leibniz Universit\"at Hannover, Institute for Gravitational
  Physics, Callinstrasse 38, 30167 Hannover, Germany}

\author{Jay Vijay Kalinani}
\address{Universit\`a di Padova, Dipartimento di Fisica e Astronomia, Via Francesco Marzolo 8, I-35131 Padova, Italy}
\address{INFN, Sezione di Padova, Via Francesco Marzolo 8, I-35131 Padova, Italy}

\author{Bruno Giacomazzo}
\address{Universit\`a di Trento, Dipartimento di Fisica, Via Sommarive 14, I-38123 Trento, Italy}
\address{INFN-TIFPA, Trento Institute for Fundamental Physics and Applications, Via Sommarive 14, I-38123 Trento, Italy}

\date{\today}

\begin{abstract}
\noindent The recent multimessenger observation of the short gamma-ray burst (SGRB) GRB\,170817A together with the gravitational wave (GW) event GW170817 provides evidence for the long-standing hypothesis associating SGRBs with binary neutron star (BNS) mergers.
The nature of the remnant object powering the SGRB, which could have been either an accreting black hole (BH) or a long-lived magnetized neutron star (NS), is, however, still uncertain.  
General relativistic magnetohydrodynamic (GRMHD) simulations of the merger process represent a powerful tool to unravel the jet launching mechanism, but so far most simulations focused the attention on a BH as the central engine, while the long-lived NS scenario remains poorly investigated. 
Here, we explore the latter by performing a GRMHD BNS merger simulation extending up to ${\sim}100$\,ms after merger, much longer than any previous simulation of this kind. This allows us to 
(i) study the emerging structure and amplification of the magnetic field and observe a clear saturation at magnetic energy $E_\mathrm{mag} \sim 10^{51}$\,erg,
(ii) follow the magnetically supported expansion of the outer layers of the remnant NS and its evolution into an ellipsoidal shape without any surrounding torus, and (iii) monitor density, magnetization, 
and velocity along the axis, observing no signs of jet formation.
We also argue that the conditions at the end of the simulation disfavor later jet formation 
on subsecond timescales if no BH is formed.
Furthermore, we examine the rotation profile of the remnant, the conversion of rotational energy associated with differential rotation, the overall energy budget of the system, and the evolution of the GW frequency spectrum.
Finally, we perform an additional simulation where we induce the collapse to a BH ${\sim}70$\,ms after merger,
in order to gain insights on the prospects for massive accretion tori in case of a late collapse.
We find that a mass around ${\sim}0.1\,M_\odot$ remains outside the horizon, which has the potential to power
a SGRB via the Blandford-Znajek mechanism when accreted.
\end{abstract}

\maketitle


\section{Introduction}
\label{sec:intro}

\noindent The first gravitational wave (GW) detection of a binary neutron star (BNS) merger by the LIGO and Virgo collaborations (event named GW170817) was accompanied by a very successful observational campaign across the entire electromagnetic spectrum, leading to crucial discoveries and demonstrating the potential of a multimessenger investigation of these unique astrophysical systems \cite{LVC-BNS,LVC-MMA,LVC-GRB}. 
In particular, the combination of the GW signal with a short gamma-ray burst (SGRB) named GRB\,170817A and the associated multiwavelength afterglow provided compelling evidence that at least a fraction of all SGRBs are produced by BNS mergers 
\cite{LVC-GRB,Goldstein2017,Savchenko2017,Troja2017,Margutti2017,Hallinan2017,Alexander2017,Mooley2018a,Lazzati2018,Lyman2018,Alexander2018,Mooley2018b,Ghirlanda2019}.

According to the widely accepted paradigm, GRBs are powered by relativistic jets, whose propagation into the interstellar medium produces the additional afterglow radiation often observed at x-ray, optical/IR, and radio wavelengths (e.g.,~\cite{Piran:2004:76,Kumar2015}). 
In full agreement with the above picture, the inferred properties of GRB\,170817A point to a jet with an energetic core of half-opening angle $\theta_\mathrm{jet,\,core}\lesssim 5^{\circ}$ and Lorentz factor $\Gamma_\mathrm{jet,\,core}\gtrsim 10$, observed $\sim\!15^{\circ}\!-\!20^{\circ}$ away from its propagation axis \cite{Mooley2018b,Ghirlanda2019}.
Moreover, the onset time of the prompt gamma-ray signal implies that this jet was launched less than $\approx\!1.74$\,s after merger \cite{LVC-MMA,LVC-GRB}.

\begin{figure*}[!ht]
  \centering
  \includegraphics[width=0.84\linewidth]{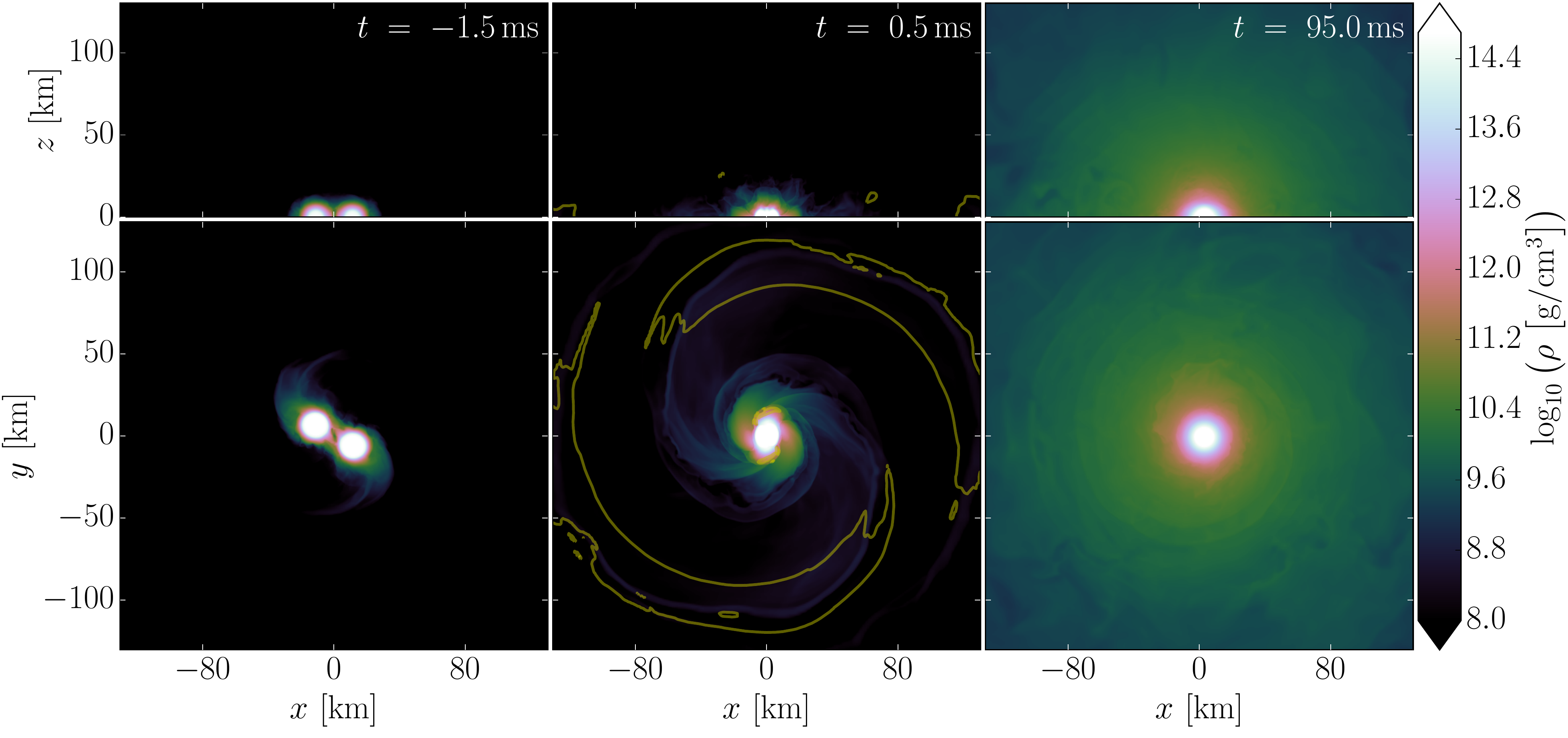}
  \caption{Rest-mass density evolution snapshots in the meridional (top panels) and equatorial (bottom panels) planes. The yellow contours indicate unbound matter ($u_t<-1$). Here $t=0$ corresponds to the merger time.} 
  \label{fig:2Drho}
\end{figure*}

While the theoretical models of this event ultimately converged to a consistent scenario, confirming that GRB\,170817A belongs to the known class of SGRBs, the nature of the central engine producing the associated jet remains an important open question. 
Most probably, the merger resulted in a metastable (hypermassive or supramassive) neutron star (NS) eventually collapsing to a black hole (BH), but the collapse could have happened at any time after merger up to several seconds later or more (e.g.,~\cite{LVC-BNS,LVC-GRB,Margalit2017,Shibata2017,Ruiz2018,LVC-rad}), i.e., before or after the SGRB jet was launched.
As a consequence, both an accreting BH (``BH-disk'' scenario) and a rapidly rotating strongly magnetized NS (``magnetar'' scenario) represent viable central engines for GRB\,170817A (see, e.g., \cite{Ciolfi2018} for a recent review). 

The question of how jets can be launched by BNS merger remnants is an unsolved problem. Current (magneto)hydrodynamics simulations suggest that magnetic fields are the dominant driver of jet formation (e.g.,~\cite{Just2016,Ruiz2016}); nevertheless, most simulations including magnetic fields focused the attention on jet formation from BH remnants (e.g.,~\cite{Kiuchi:2014:41502,Ruiz2016,Kawamura:2016:064012}), leaving the case of a massive NS central engine essentially unexplored. 
Recently, Ciolfi {\it et al.}~(2017) \cite{Ciolfi2017} started the first systematic investigation of magnetized BNS mergers leading to long-lived NS remnants\footnote{Hereafter, long-lived refers to remnants with lifetime $\gg\!100$\,ms.} and addressed, among various aspects, the potential to form a jet without a BH. While no favorable indications were found, this study showed that the relevant timescales for jet formation can be significantly longer than in the BH case, suggesting that a much longer postmerger evolution would be necessary to obtain more conclusive results.

Here, we present a new general relativistic magnetohydrodynamics (GRMHD) simulation of a BNS merger forming a long-lived NS, where we follow the evolution up to 95\,ms after merger. This represents to date the longest simulation of this kind (previously, the longest postmerger evolution extended up to 45\,ms \cite{Ciolfi2017}). 
We study the amplification and rearrangement of magnetic fields, observing a clear saturation of the magnetic energy growth and getting hints that a global field structure might be forming.
Next, we monitor the outflows and the magnetic-to-fluid pressure ratio along the orbital axis to look for signs of jet formation. 
Further, we investigate in detail the energy budget of the system, in particular the evolution of rotational energy.  
Finally, we complete our analysis by looking at the structure and the rotational profile of the remnant NS as well as the GW emission.

In addition to the above simulation, we perform a second one where we induce the collapse to a BH $\sim\!70$\,ms after merger. This allows us to study the effects of delayed BH formation. Based on the mass and magnetic field strength remaining outside the horizon, we argue that the resulting system might be able to produce a successful SGRB jet compatible with GRB\,170817A. 

The paper is organized as follows. Section~\ref{sec:dynamics} presents the BNS model, the general dynamics, and matter outflows. Section~\ref{sec:mag} focuses on the evolution, amplification, and emerging geometrical structure of the magnetic field. Section~\ref{sec:sgrb} is devoted to the possibility of jet formation and the connection with SGRBs. In Sec.~\ref{sec:remnant}, we turn to the remnant NS structure and rotation profile evolution.
The GW emission is discussed in Sec.~\ref{sec:oscil}. 
We summarize the results and add concluding remarks in Sec.~\ref{sec:conclusion}.


\section{Merger and postmerger dynamics}
\label{sec:dynamics}

\noindent Our simulation evolves a $1.35\!-\!1.35\,M_\odot$ BNS system obeying the APR4 equation of state (EOS) \cite{Akmal:1998:1804}, endowed with an initial dipolar magnetic field confined to the interior of the stars. 
The physical and numerical setup is identical to the equal-mass APR4 model discussed in \cite{Ciolfi2017}, except
for one important difference: here we impose a very large initial magnetic energy of $E_\mathrm{mag}\simeq 1.6\times 10^{48}$\,erg, corresponding to a maximum magnetic field strength of $B_\mathrm{max}\simeq 10^{16}$\,G, whereas in \cite{Ciolfi2017} we chose $B_\mathrm{max}\simeq 3\times10^{15}$\,G. 
 
We have chosen this magnetic field strength in order to emphasize magnetic field effects and obtain more favorable conditions for jet formation (Sec.~\ref{sec:sgrb}) within a shorter time span. Although such a strong field in merging NS is unlikely to occur in
nature, it compensates for the fact that our resolution is not sufficient to fully account for the main magnetic field amplification mechanisms (a common limitation for current numerical relativity simulations).
In reality, one might reach the same field strength at a given time after merger when starting with weaker fields.
For further discussion, we refer to  \cite{Kiuchi2015,Kiuchi:2018,Giacomazzo:2015}.

\begin{figure}[!ht]
  \centering
  \includegraphics[width=0.94\linewidth]{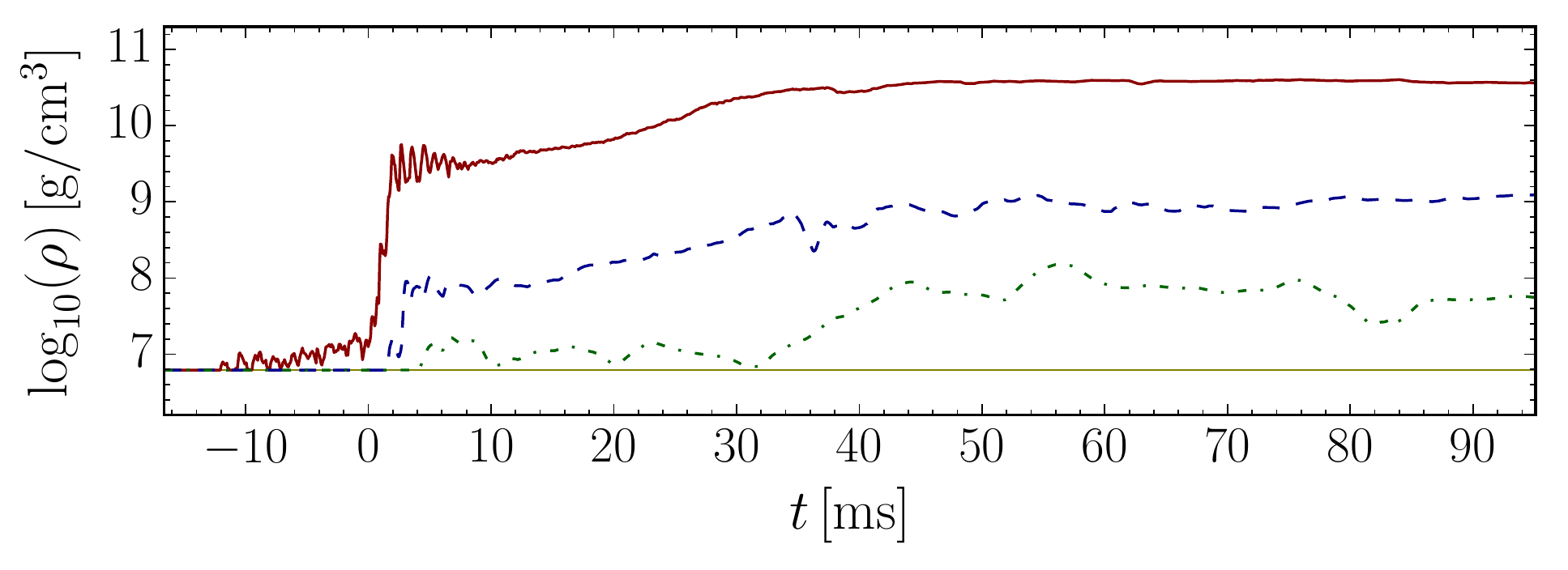}
  \includegraphics[width=0.94\linewidth]{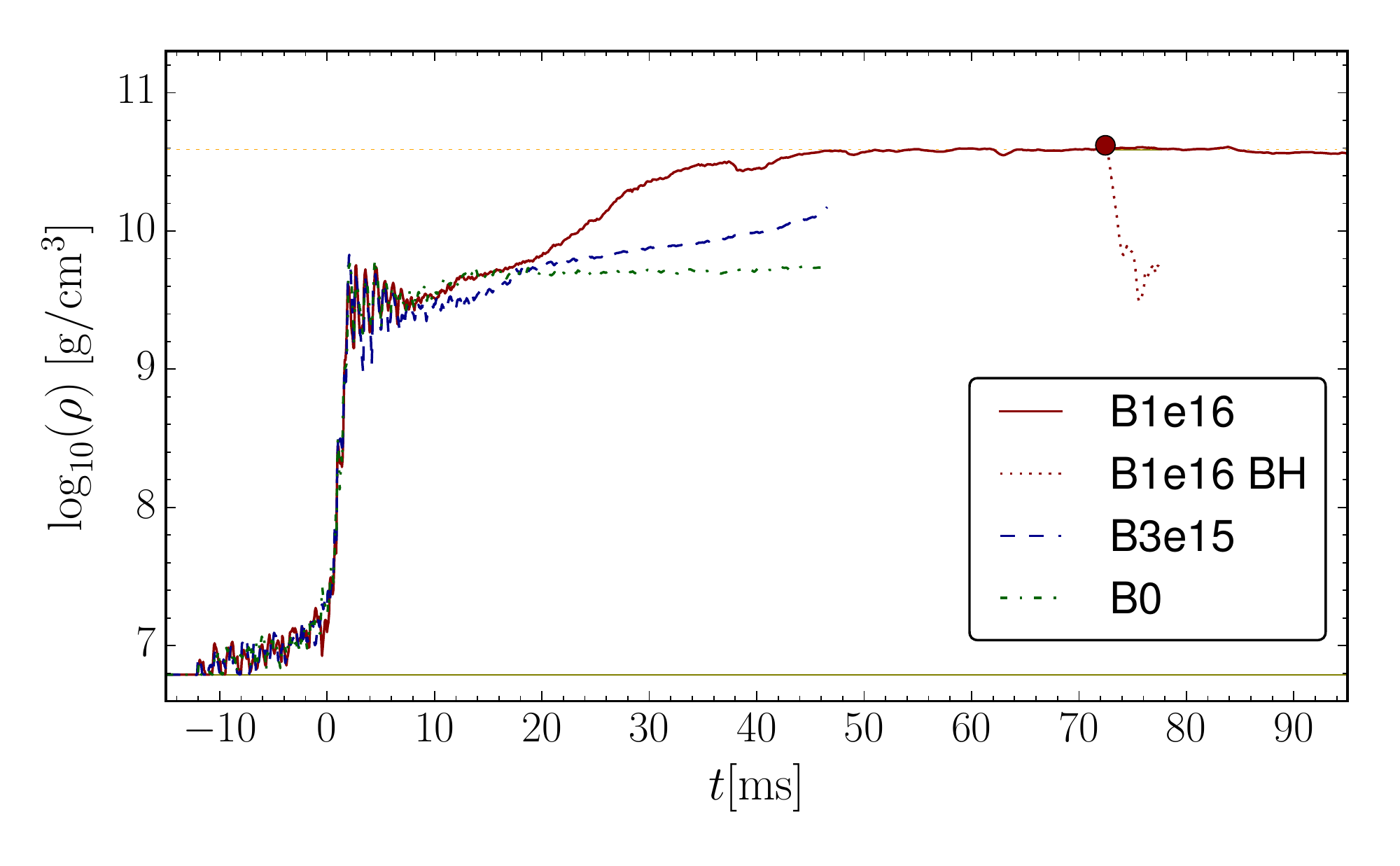}
  \caption{{\it Top:} Rest-mass density along the orbital axis averaged in the distance ranges $30-50$\,km, $150-200$\,km, and $400-500$\,km (continuous, dashed, and dot-dashed lines). The horizontal line marks the density of the artificial atmosphere.
{\it Bottom:} Rest-mass density along the orbital axis averaged in the distance range $30-50$\,km for the present simulation with and without collapse (``B1e16 BH'' and ``B1e16'') and for the lower and zero magnetic field models in \cite{Ciolfi2017} (``B3e15'' and ``B0'').} 
  \label{fig:rhofunnel}
\end{figure}

Figure~\ref{fig:2Drho} shows meridional and equatorial snapshots of the rest-mass density from the merger time to the end of the simulation.
The merger results in a long-lived supramassive NS remnant,\footnote{The supramassive regime for the given EOS corresponds to a remnant NS mass above the maximum mass for a nonrotating NS, but below the maximum mass for a uniformly rotating NS.} reaching a quasistationary and highly axisymmetric configuration in $\sim\!50$\,ms. 
In the following $\sim\!45$\,ms of evolution, the density distribution remains fairly constant up to very large radii ($\sim\!400$\,km), in particular along the remnant spin axis [Fig.~\ref{fig:rhofunnel}(a)]. 
For radial distances between 50 and 400\,km, density distribution is also nearly isotropic and follows a radial profile $\propto\!1/r^2$ up to $\sim\!150$\,km becoming slightly steeper for larger radii.
Below $r\approx50$\,km, the remnant deviates from spherical symmetry because of a torus-shaped outer envelope (Sec.~\ref{sec:remnant} and Fig.~\ref{fig:remnant}), while for $r>400$\,km the density distribution gradually acquires a polar angle dependence (see below). 

Compared to lower magnetizations, the remnant is surrounded by a much denser environment. For instance, at 30\,km distance along the spin axis the rest-mass density is $\sim4\times10^{10}$\,g/cm$^3$, about 1 order of magnitude higher than the nonmagnetized case [Fig.~\ref{fig:rhofunnel}(b)]. This indicates that the postmerger baryon-loaded wind polluting the environment is primarily induced by magnetic pressure at this magnetization level (see also Fig.~23 in \cite{Ciolfi2017}).

\begin{figure*}[!ht]
  \centering
  \includegraphics[width=0.47\linewidth]{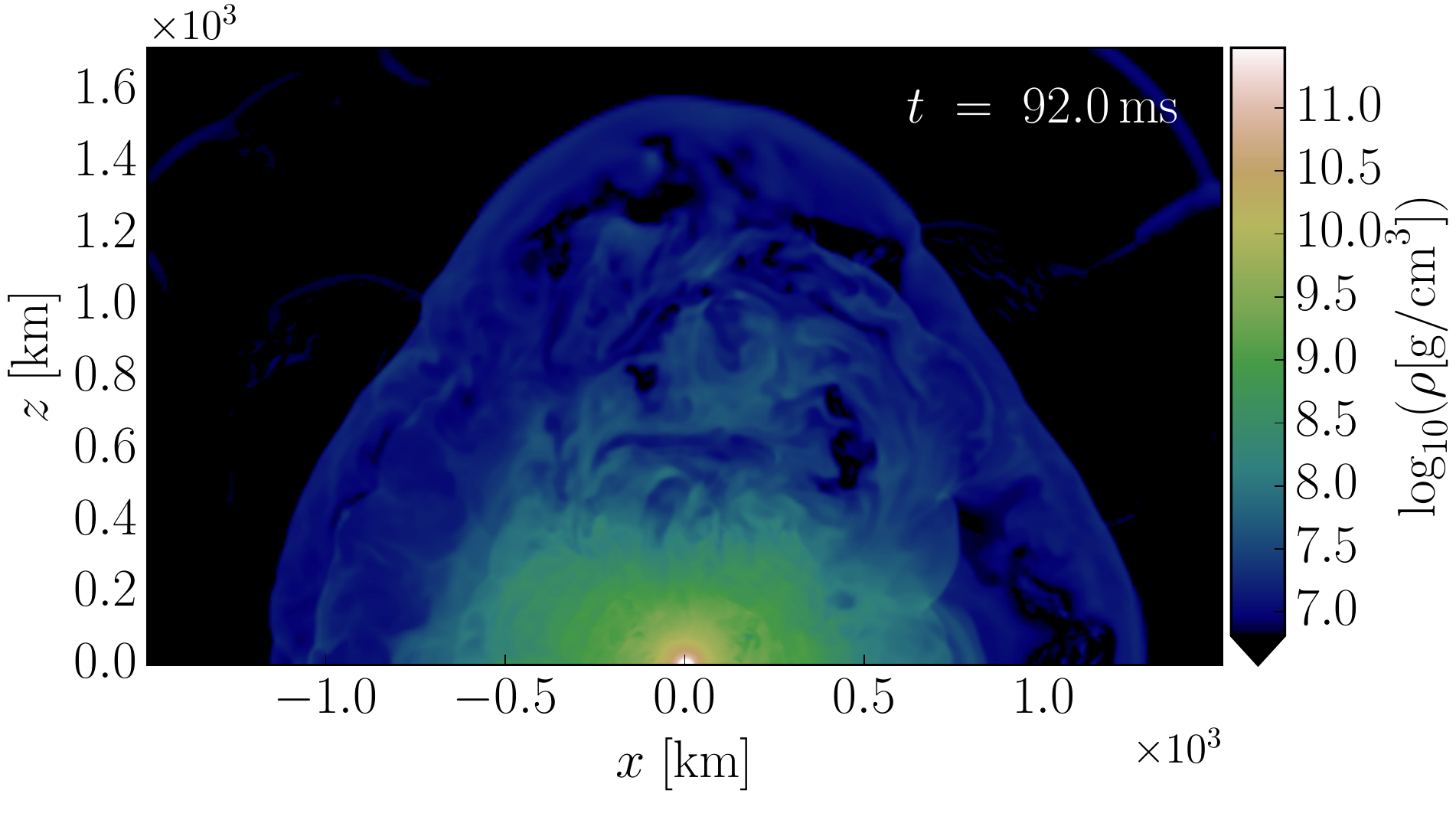}
\hspace{0.5cm}
  \includegraphics[width=0.485\linewidth]{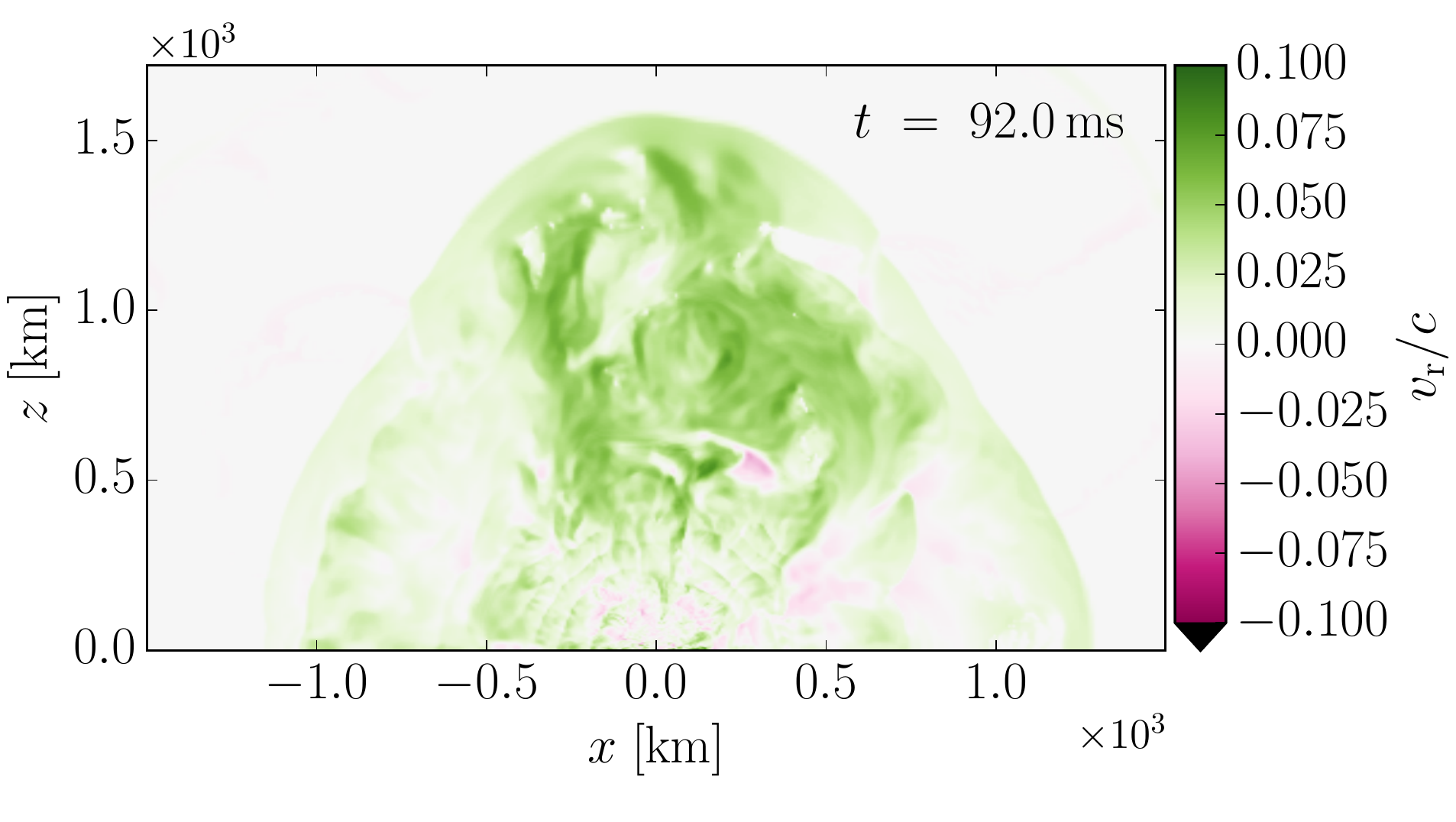}
  \caption{{\it Left:} Large scale meridional view of the rest-mass density toward the end of the simulation. {\it Right:} Radial velocity of the outflowing material (positive values indicate motion directed outwards).} 
  \label{fig:rho_vel}
\end{figure*}

At larger scales a magnetized outflow emerges, composed by a faster polar component (within $\sim\!40^{\circ}$ from the remnant spin axis) moving at a radial velocity up to 0.1\,c and a slower isotropic component with radial velocity below 0.05\,c (Fig.~\ref{fig:rho_vel}),
Across a spherical surface of radius $r=300$\,km, we estimate a very high mass flow rate reaching $\dot{M}\!\sim3\,M_{\odot}/$s around 40\,ms after merger and decaying to $\sim\!1\,M_{\odot}/$s toward the end of the simulation. The corresponding cumulative mass outflow is $\sim\!0.1\,M_{\odot}$, excluding the early contribution from the tidal and shock-driven dynamical ejecta ($\simeq\!0.015\,M_{\odot}$). 
A significant fraction of this matter might become unbound at large distances, leading to a very massive ejecta component. We note that magnetized postmerger outflows like this represent one of the leading candidates to explain the early (or ``blue'') part of the radioactively powered kilonova signal observed in association with GW170817 (e.g.,~\cite{Metzger2018}).
A direct comparison with the fiducial model in \cite{Ciolfi2017} suggests a strong dependence of $\dot{M}$ on the magnetic energy, i.e.,~$\dot{M}\propto(E_\mathrm{mag})^{\,a}$ with $1\!\lesssim\!a\!\lesssim2$ (only valid for magnetically dominated mass ejection, i.e.,~above a certain threshold magnetization).

Note that we are imposing an artificial floor density of $\rho^*\simeq6.3\times10^{6}$\,g/cm$^3$ (as in \cite{Ciolfi2017}), which corresponds to $\simeq0.4$\,(13)$\times10^{-3}\,M_{\odot}$ within a sphere of radius 300\,(1000)\,km.
While the outflow across $r=300$\,km should not be significantly affected, the radial velocity at larger distances is likely reduced by the presence of the floor density.

\begin{figure}[!ht]
  \centering
  \includegraphics[width=0.94\linewidth]{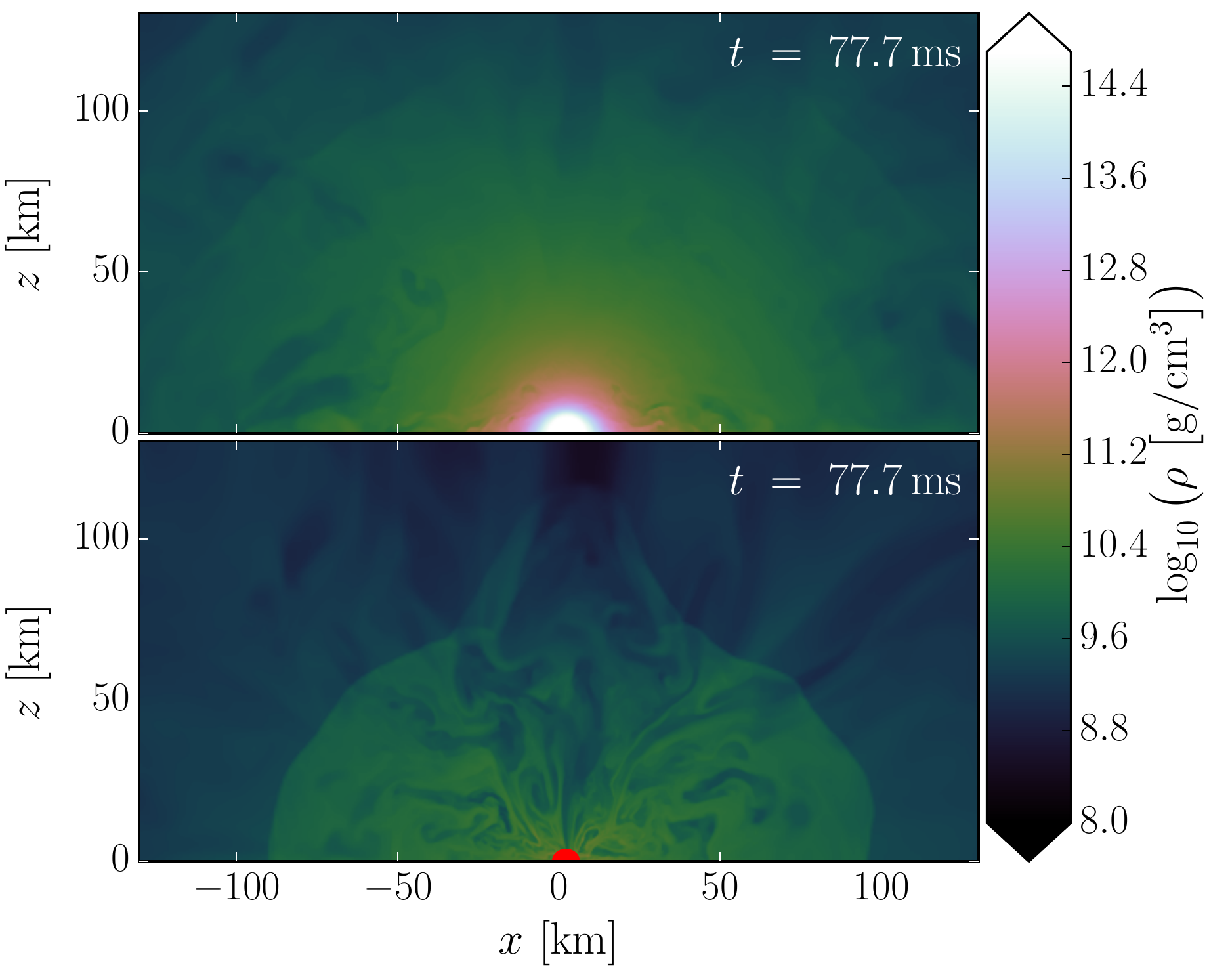}
  \caption{Meridional view of rest-mass density for the noncollapsing (top) and collapsing (bottom) cases. The filled region in red indicates the black hole horizon.} 
  \label{fig:LLvsBH}
\end{figure}

In order to explore how the collapse to a BH would affect the surrounding environment, we performed a second simulation starting from 72\,ms after merger where we induced the collapse by modifying the EOS parameters at supranuclear densities.\footnote{We changed the piecewise-polytropic parameters at $\rho>2\times10^{14}$\,g/cm$^3$ to obtain a less stiff EOS in the core (in particular, we substituted the parameter values of APR4 with the ones of H4 \cite{Ciolfi2017}); note that the portion of the NS remnant affected by this modification is entirely swallowed within the BH horizon.}
As a result, a BH is suddenly formed and we followed the evolution for another $\sim\!5$\,ms (Fig.~\ref{fig:LLvsBH}). At the end, the final BH mass and dimensionless spin are $M_\mathrm{BH}\simeq2.5\,M_{\odot}$ and $\chi\equiv Jc/GM^2\simeq0.5$. Around it, an accretion torus of $M_\mathrm{disk}\simeq0.1\,M_{\odot}$ is forming, while along the spin axis a low density funnel is created (at $r=30$\,km, for instance, rest-mass density is 1 order of magnitude smaller than the noncollapsing case; see Fig.~\ref{fig:LLvsBH}).

\begin{figure}[!ht]
  \centering
  \includegraphics[width=0.98\linewidth]{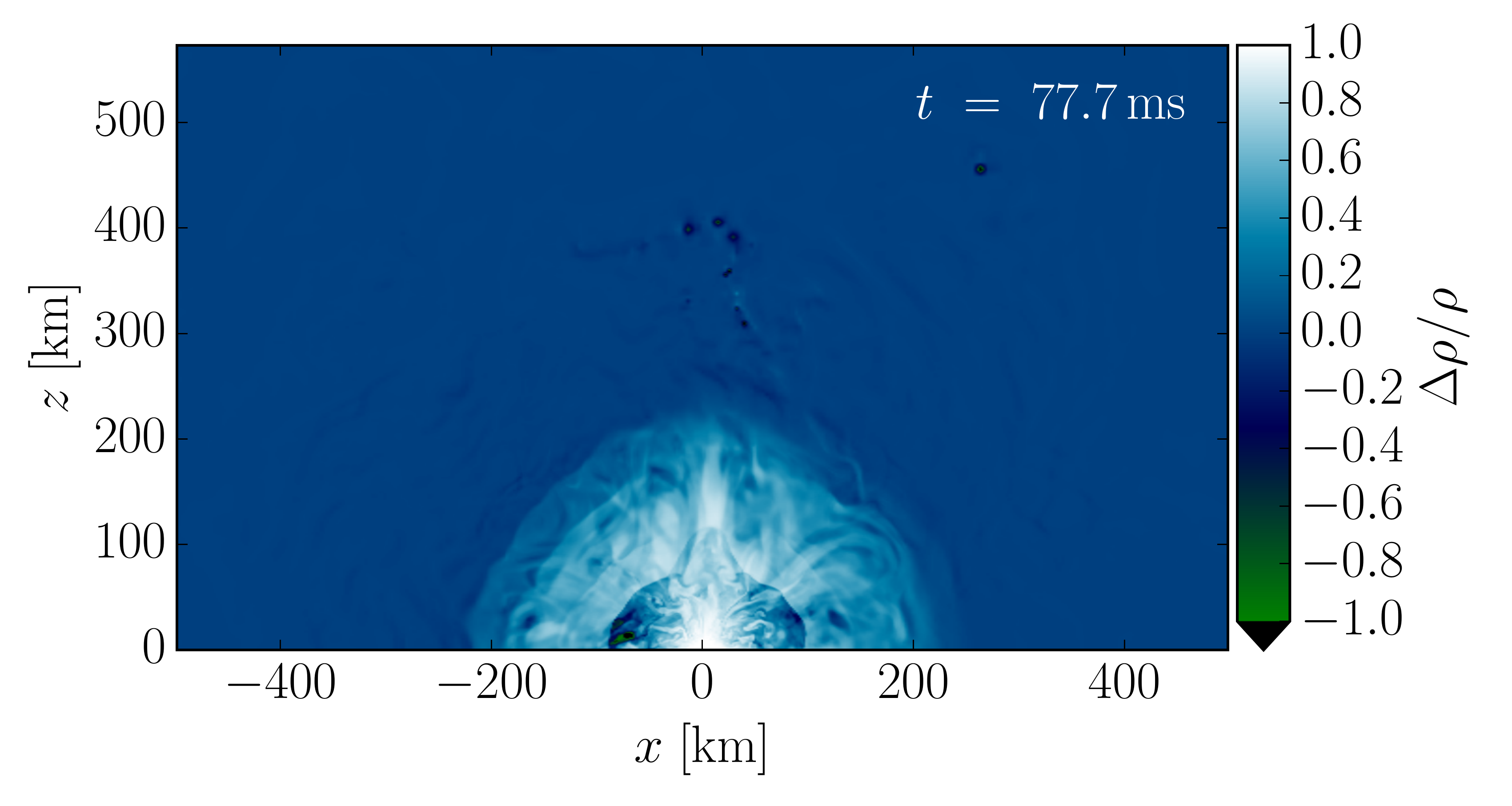}
  \caption{Meridional view of the relative difference in rest-mass density between the noncollapsing and collapsing cases.}
  \label{fig:rhodiff}
\end{figure}

Figure~\ref{fig:rhodiff} shows the relative difference in rest-mass density between the collapsing and noncollapsing cases. Only the matter within $\sim\!200$\,km is affected, while the rest of the system is not directly influenced by BH formation. 


\section{Magnetic fields}
\label{sec:mag}

\begin{figure}[!ht]
  \centering
  \includegraphics[width=0.88\linewidth]{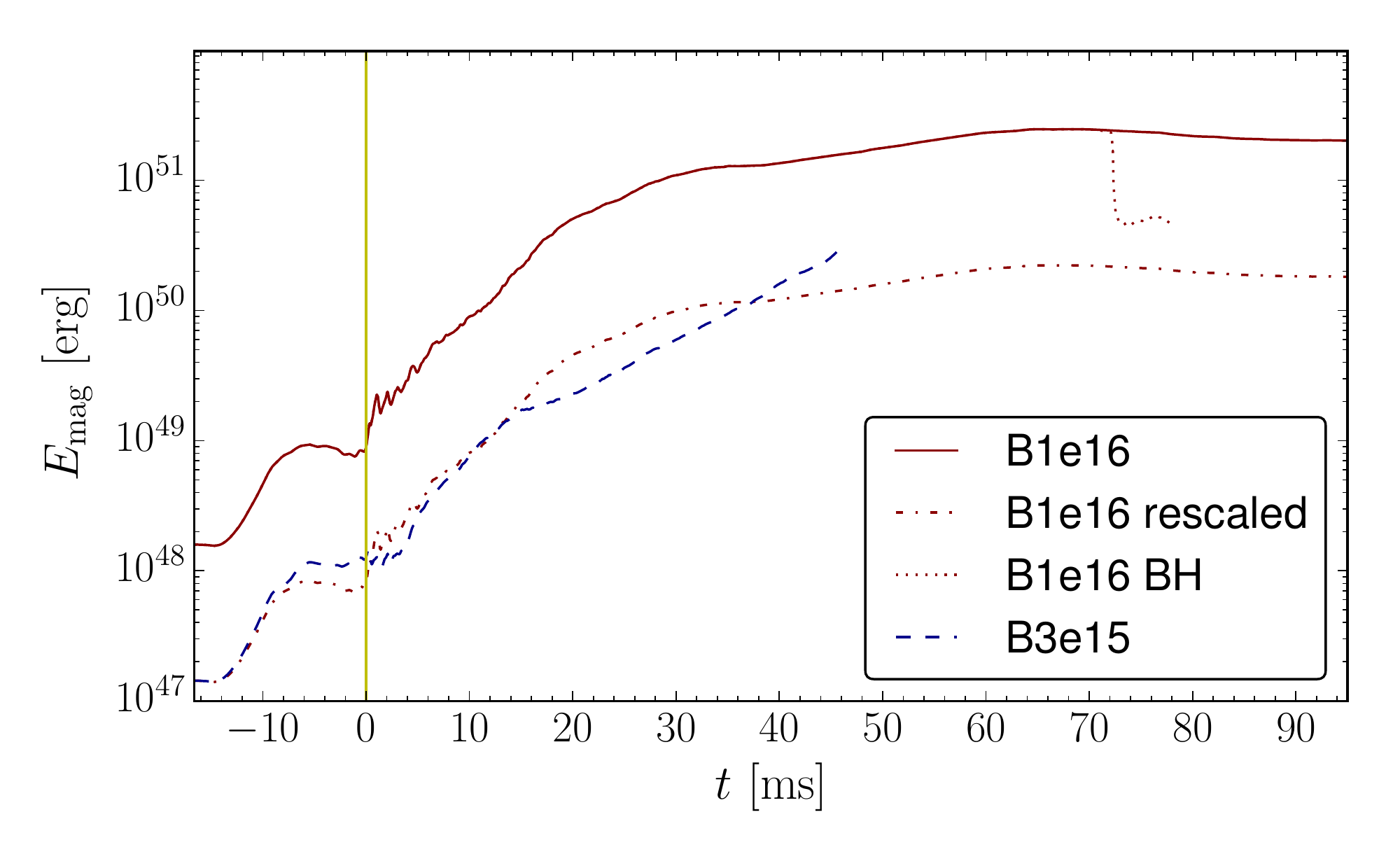}
  \includegraphics[width=0.877\linewidth]{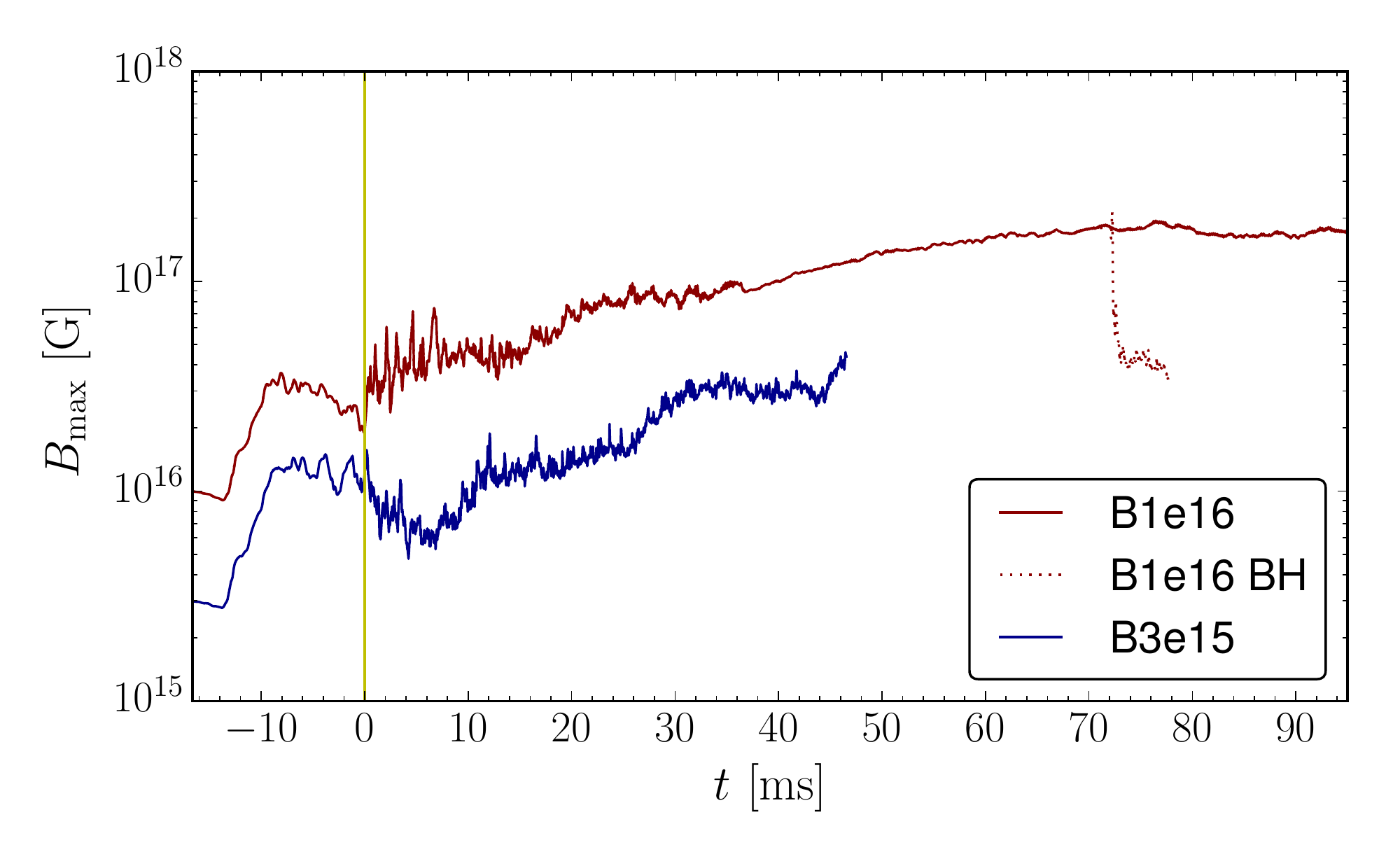}
  \caption{Evolution of magnetic energy (top) and maximum magnetic field strength (bottom) for the present simulation with and without collapse (``B1e16 BH'' and ``B1e16'') and for the lower magnetic field model in \cite{Ciolfi2017} (``B3e15''). Vertical line marks the time of merger. The top panel shows also the magnetic energy of model ``B1e16'' rescaled to match the initial value of model ``B3e15''.} 
  \label{fig:Emag-Bmax}
\end{figure}

\noindent We now turn to discuss the magnetic field evolution.
As shown in Fig.~\ref{fig:Emag-Bmax}, magnetic fields undergo different stages of amplification up to a saturation achieved around 50\,ms after merger. 

The premerger amplification is likely associated with fluid flows developing inside the two NSs during the late inspiral, possibly induced by the time-varying tidal deformations \cite{Ciolfi2017}. However, nonphysical effects due to the initial data or the simulation setup cannot be excluded, and the origin of this amplification remains a matter of debate \cite{Ciolfi2017}. 

At merger, magnetic field amplification is dominated by the Kelvin-Helmholtz (KH) instability (e.g.,~\cite{Kiuchi2015,Kiuchi:2018}), acting in the shear layer separating the two NS cores and increasing the magnetic energy by almost 1 order of magnitude in a few ms.
Then, the KH instability is gradually substituted by magnetic winding and the magnetorotational instability (MRI) as the dominant source of amplification \cite{Balbus:1991,Duez2006a,Siegel:2013:121302}. 
At the final saturation level, magnetic energy is $E_\mathrm{mag}\simeq 2\times 10^{51}$\,erg and the magnetic field strength  reaches a maximum of $B_\mathrm{max}\simeq 10^{17}$\,G.

\begin{figure}[!ht]
  \centering
  \includegraphics[width=0.90\linewidth]{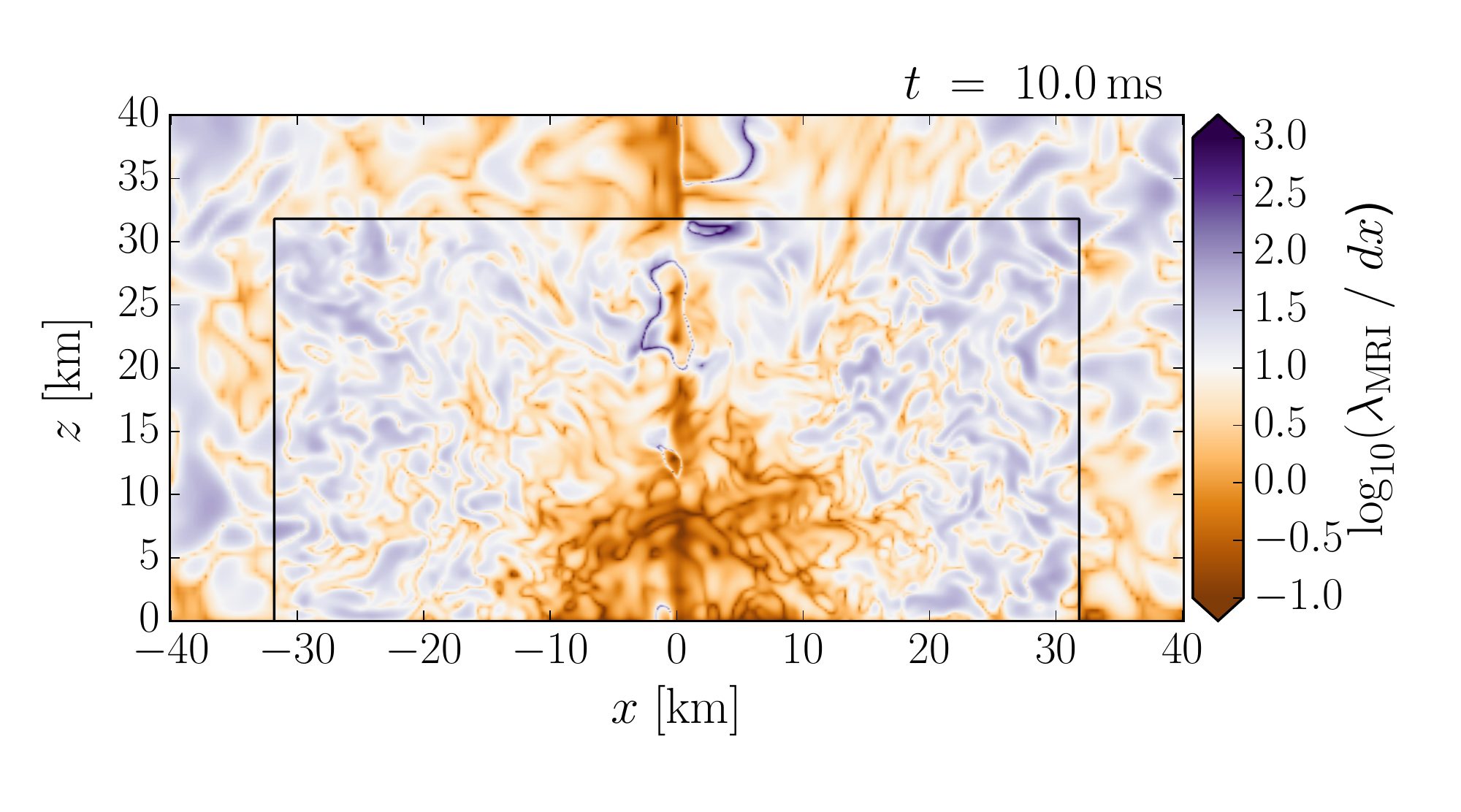}
  \includegraphics[width=0.90\linewidth]{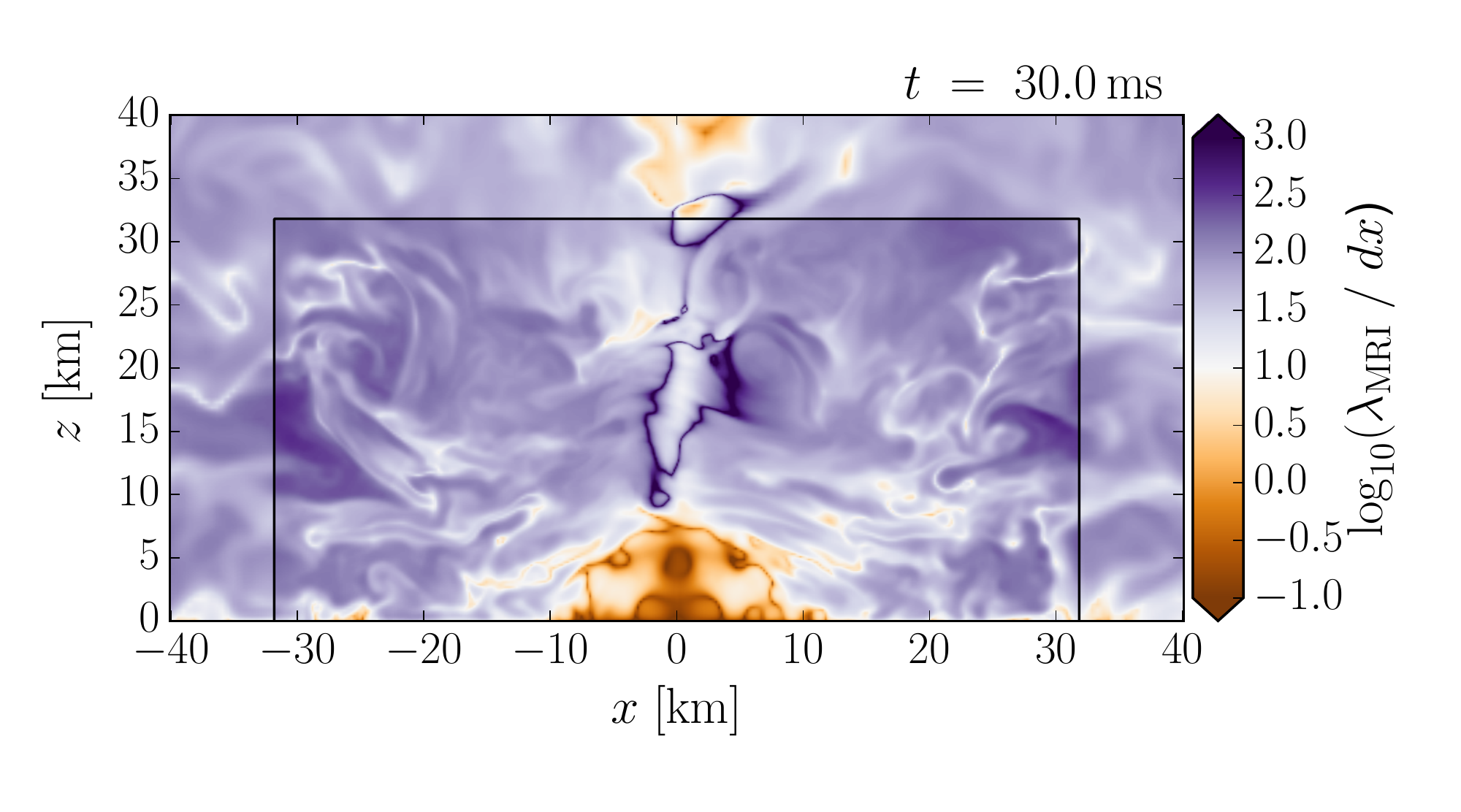}
  \caption{Meridional view of the number of grid points covering 
  $\lambda_\mathrm{MRI}$ (see text) 10 and 30\,ms after merger. 
  Solid black line marks the transition from the central finest refinement 
  level and the second-finest one, where the latter has a factor of 2 
  larger grid spacing $dx$.} 
  \label{fig:MRI}
\end{figure}

As demonstrated in \cite{Kiuchi2015,Kiuchi:2018}, the numerical resolution employed here (minimum grid spacing of $dx\approx220$\,m) is insufficient to fully capture the small-scale amplification due to the KH instability, and we therefore expect 
the magnetic energy to increase even faster.
On the other hand, as the magnetic field strength keeps increasing, the wavelength of the fastest growing MRI mode becomes easier to resolve. In Fig.~\ref{fig:MRI}, we estimate the latter according to $\lambda_\mathrm{MRI} \approx (2\pi/\Omega) \times B/\sqrt{4\pi\rho}$ (where $\Omega$, $B$, and $\rho$ are angular velocity, magnetic field strength, and rest-mass density) and compare with the grid resolution at 10 and 30\,ms after merger. We find that at 30\,ms $\lambda_\mathrm{MRI}$ is covered everywhere by $10-1000$ grid points, except for the inner region enclosed in a 10\,km radius. Since the MRI becomes effective when $\lambda_\mathrm{MRI}/dx\gtrsim10$, this result suggests that for $t\gtrsim30$\,ms (where $t=0$ is the merger time) and $r_\mathrm{cyl}\gtrsim10$\,km the dominant magnetic field amplification mechanism is well resolved.
Note that the MRI is active in the presence of differential rotation, only where the angular velocity profile is decreasing with the distance from the spin axis. As will be discussed in Sec.~\ref{sec:remnant}, this is not the case in the region $r<10$\,km, which would therefore not be subject to the MRI anyway (see Fig.~\ref{fig:rot_prof_evol}). We caution, however, that the rotation profile is likely not fully converged (compare also Fig.~3 in \cite{Kiuchi:2018}).
\begin{figure*}[!ht]
  \centering
  \includegraphics[width=0.85\linewidth]{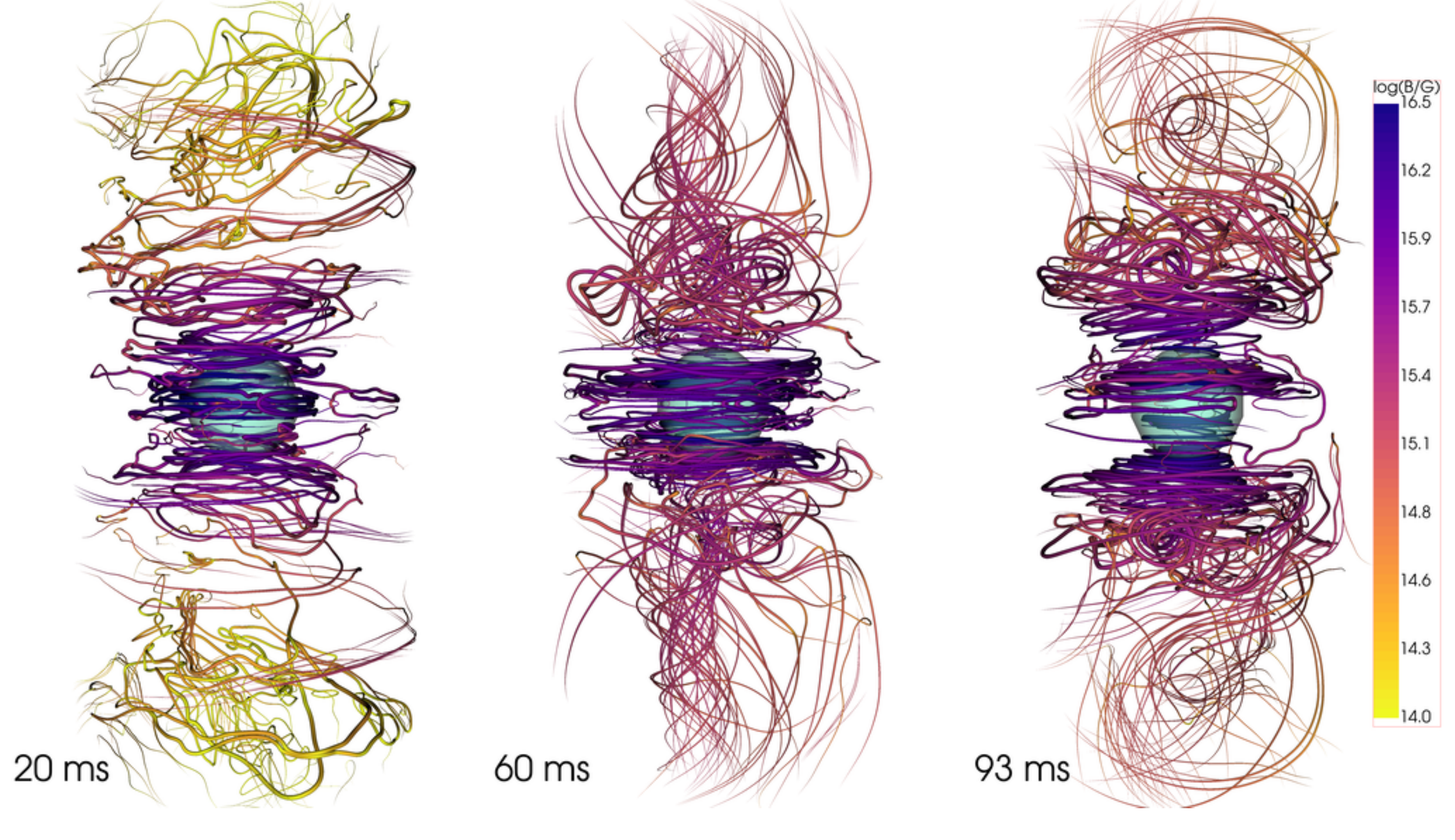}
  \caption{Magnetic field structure 20, 60, and 93\,ms after merger. To give a scale reference, we added a (cyan) sphere of 10\,km radius placed at the remnant center of mass. Field line colors indicate the magnetic field strength.}
  \label{fig:3D-B}
\end{figure*}
\begin{figure}[!ht]
  \centering
  \includegraphics[width=0.94\linewidth]{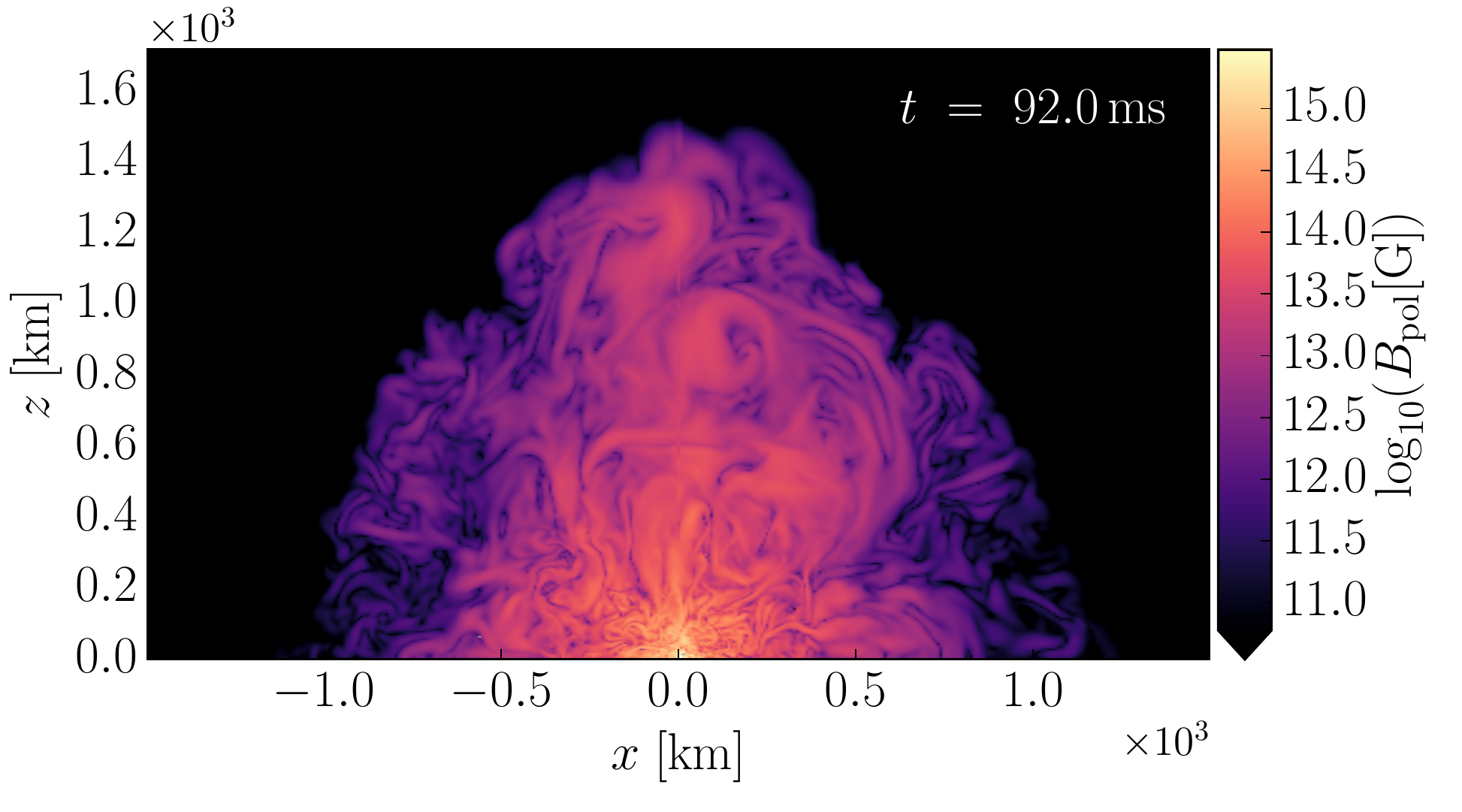}
  \includegraphics[width=0.94\linewidth]{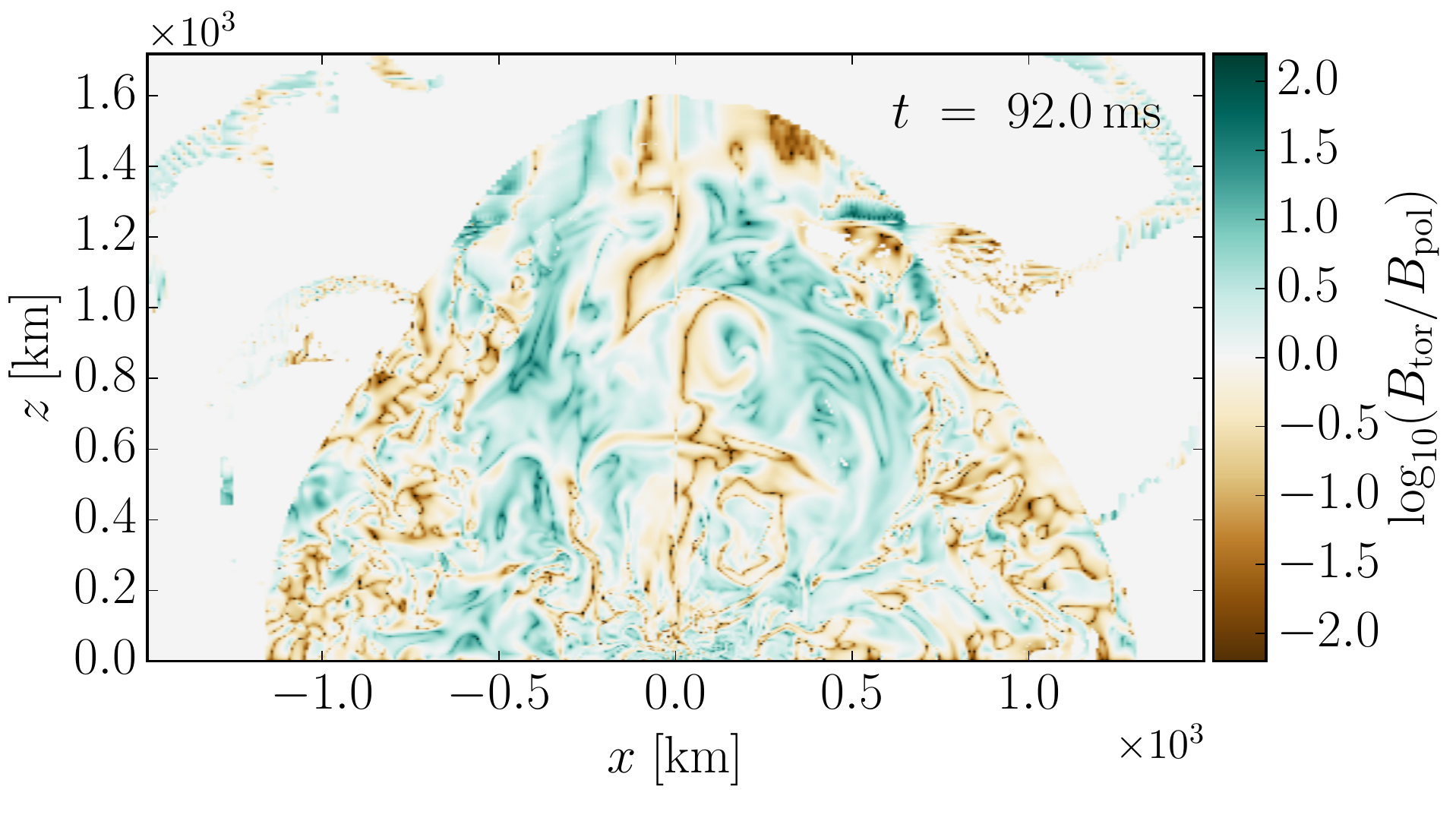}
  \caption{Large scale meridional view of the poloidal magnetic field strength (top) and the toroidal-to-poloidal field strength ratio (bottom). To help visualization, the latter is set to zero where the rest-mass density is close to the artificial floor value ($<1.5\times\rho^*$).}
  \label{fig:2D-PolTor}
\end{figure}

In Fig.~\ref{fig:Emag-Bmax} we also show the result obtained in \cite{Ciolfi2017} with a lower initial magnetization. 
To ease comparison, we also show the result of the present simulation rescaled to the same initial $E_\mathrm{mag}$.
As one can see, the initial evolution is very similar. During the period $20$--$40$ ms after merger, saturation
becomes evident for the present simulation, while the results from \cite{Ciolfi2017} show no sign of saturation.
Instead, the slope is comparable to the present case at the same magnetic energy.
The simulation with lower field is not long enough to determine when saturation would begin,  
reaching magnetic energies just before the onset of saturation in the present simulation.  
However, it seems plausible that the onset of saturation depends mostly on the magnetic field energy.
Based on this, we do not expect that the system could evolve toward $E_\mathrm{mag}$ much larger than $\sim\!10^{51}$\,erg, also in agreement with, e.g., \cite{Kiuchi2015,Giacomazzo:2015}.
Therefore, the magnetization obtained in the present long simulation should represent an indicative upper limit to what a realistic BNS merger can produce (for the given mass and EOS).

In case of a collapse to a BH, a large fraction of magnetic energy is expected to be rapidly accreted and thus removed from the system. The top panel of Fig.~\ref{fig:Emag-Bmax} shows the corresponding energy jump for the collapsing case we simulated (BH formation at 72\,ms after merger), revealing that only $\sim\!20$\% of magnetic energy survives. 
This limits the energy reservoir that could produce a magnetic explosion, as envisaged by some models (e.g.,~\cite{Nathanail2019}). 
After the collapse, however, further magnetic field amplification is possible within the accretion disk surrounding the BH. 

Figure~\ref{fig:3D-B} shows the magnetic field line structure close to the remnant core at 20, 60, and 93\,ms after merger. We employ the same procedure of \cite{Kawamura:2016:064012} to show the magnetic field lines that are (on average) strongest along
cones of constant angle to the rotation axis. 
As expected, toroidal fields in the equatorial region are strongly amplified. 
Careful inspection also reveals a narrow ordered helical field embedded in a more irregular field. 
It is generated between 20 to 60\,ms after merger 
around the remnant spin axis. The poloidal component is dominant, 
and the field strength grows from $\sim\!10^{14}$ to more than $10^{15}$\,G. 
At later times, the helical structure is no longer visible, substituted by a more disordered field line distribution along the axis. Toward the end of the simulation, the strongest field lines are predominantly toroidal up to 20\,km from the remnant center. 

\begin{figure*}[!ht]
  \centering
  \includegraphics[width=0.85\linewidth]{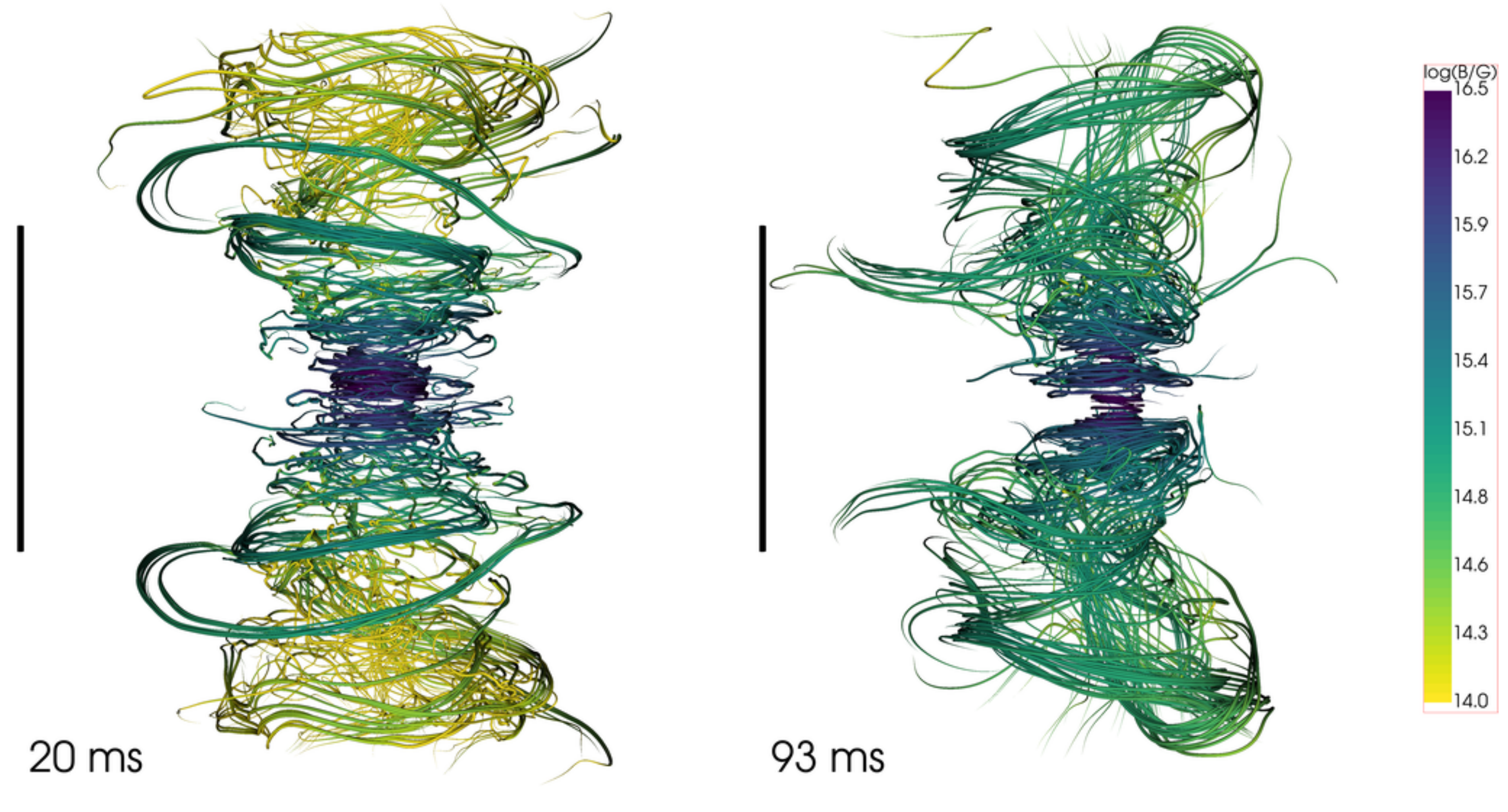}
  \caption{Same as Fig.~\ref{fig:3D-B} at a larger spatial scale. The vertical bar corresponds to 100\,km.}
  \label{fig:3D-B_large}
\end{figure*}

A larger scale visualization (Fig.~\ref{fig:3D-B_large}) shows magnetic field lines of strength $\gtrsim\!10^{15}$\,G and mixed toroidal-poloidal character, which wind around the axis, roughly along a wide cone. At the end of the simulation,
those lines loop around when reaching distances around $100$ km from the remnant. 
The $\sim\!100$\,km extension of these field lines suggests that a global magnetic field structure might be forming.
Note that we imposed initial premerger magnetic fields confined within the two NSs, with no field extending to the exterior. Therefore, the emergence of a large scale field is entirely due to the dynamical and geometrical properties of the remnant.
\begin{figure}[!ht]
  \centering
  \includegraphics[width=0.94\linewidth]{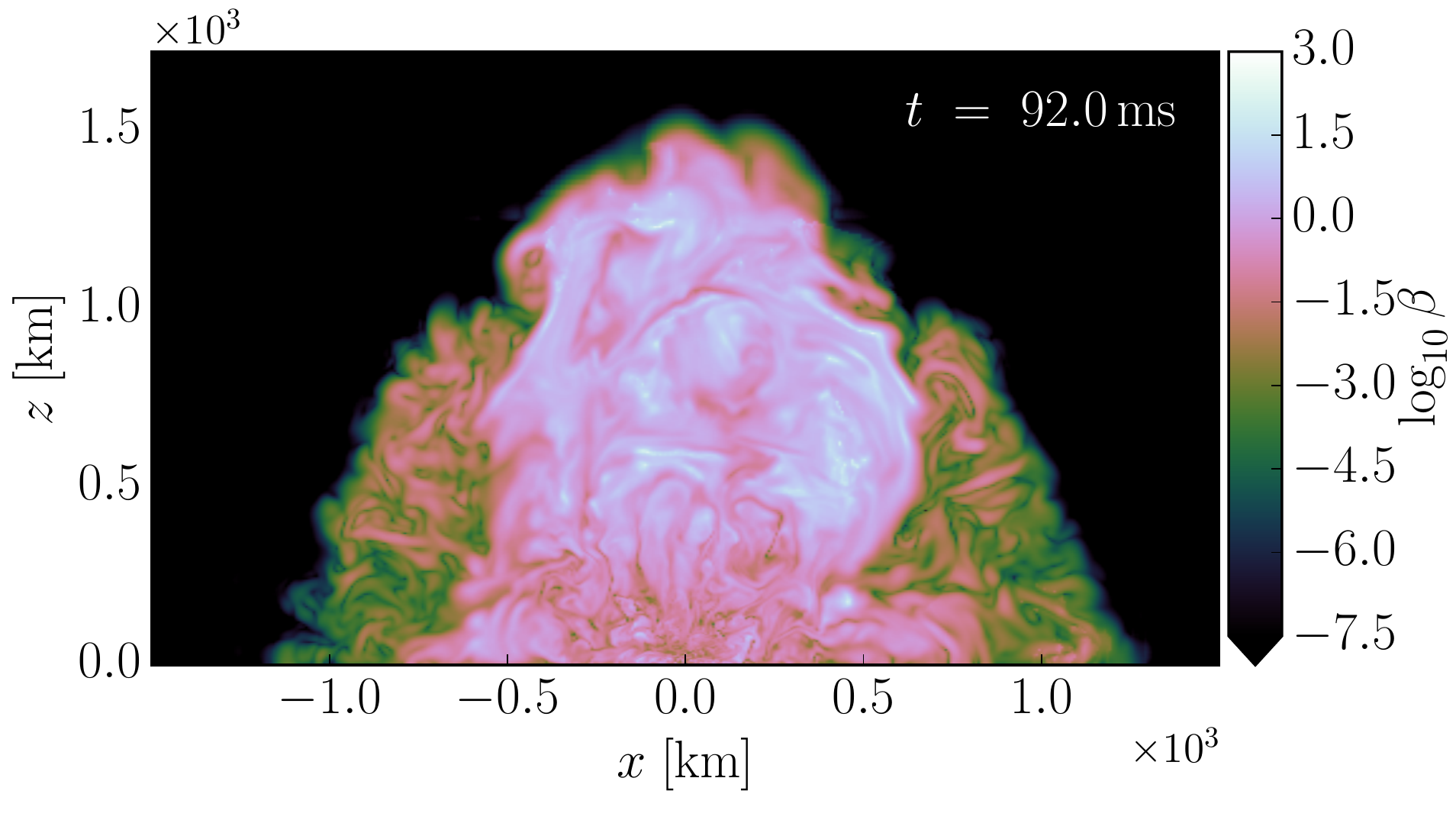}
  \caption{Large scale meridional view of the magnetic-to-fluid pressure ratio (see text).}
  \label{fig:2D-Beta}
\end{figure}

Looking  at the largest scales (Fig.~\ref{fig:2D-PolTor}), we find that the magnetic field strength follows the more turbulent distribution of the outflowing material (i.e.,~the baryon-loaded wind). 
Toroidal and poloidal field components are comparable, with the toroidal one being slightly stronger in the largest part of the volume. 
Along the spin axis, the poloidal field strength is still almost as strong as $10^{14}$\,G up to radii above $1000$\,km.


\section{Short gamma-ray bursts}
\label{sec:sgrb}

\noindent One main goal of the present work is to investigate whether a long-lived NS remnant would be able to launch a jet compatible with a SGRB (and in particular GRB\,170817A).
By the end of our simulation, we do not obtain any direct sign of jet formation. 
Along the polar direction, we observe a noncollimated, nonrelativistic, and rather dense magnetized outflow, with magnetic-to-fluid pressure ratio $\beta\equiv b^2/2p$ between $0.1$ and 10 in most of the occupied volume (Fig.~\ref{fig:2D-Beta}; here $b^2\equiv b^\mu b_\mu$ and $b^\mu$ is the 4-vector of the magnetic field as measured by the comoving observer \cite{Giacomazzo:2007:235}). An incipient jet would require a much faster, more magnetically dominated, and lower density outflow (e.g.,~\cite{Ruiz2016}). 
The baryon-loaded wind present at 90\,ms after merger is also too heavy to be accelerated up to Lorentz factors $\Gamma>10$, as necessary for SGRB jets.\footnote{It would correspond to a kinetic energy larger than $\sim\!10^{54}$\,erg.}
Instead, it represents a potential obstacle for any jet launched by the remnant at later times, requiring additional energy to drill through it. 

If an incipient jet is actually produced at a later time along the spin axis, the baryon density would have to be very small at the launching radius. 
Up to $\sim\!400$\,km distance, the density is too high ($\gtrsim10^8$\,g/cm$^3$) and no appreciable decrease is observed over $\sim\!100$\,ms (Fig.~\ref{fig:rhofunnel}), implying a timescale for jet formation $\gg\!100$\,ms (if any). 
Larger jet launching radii appear unlikely due to the diffuse nature of the matter distribution.

\begin{figure}[!t]
  \centering
   \includegraphics[width=0.999\linewidth]{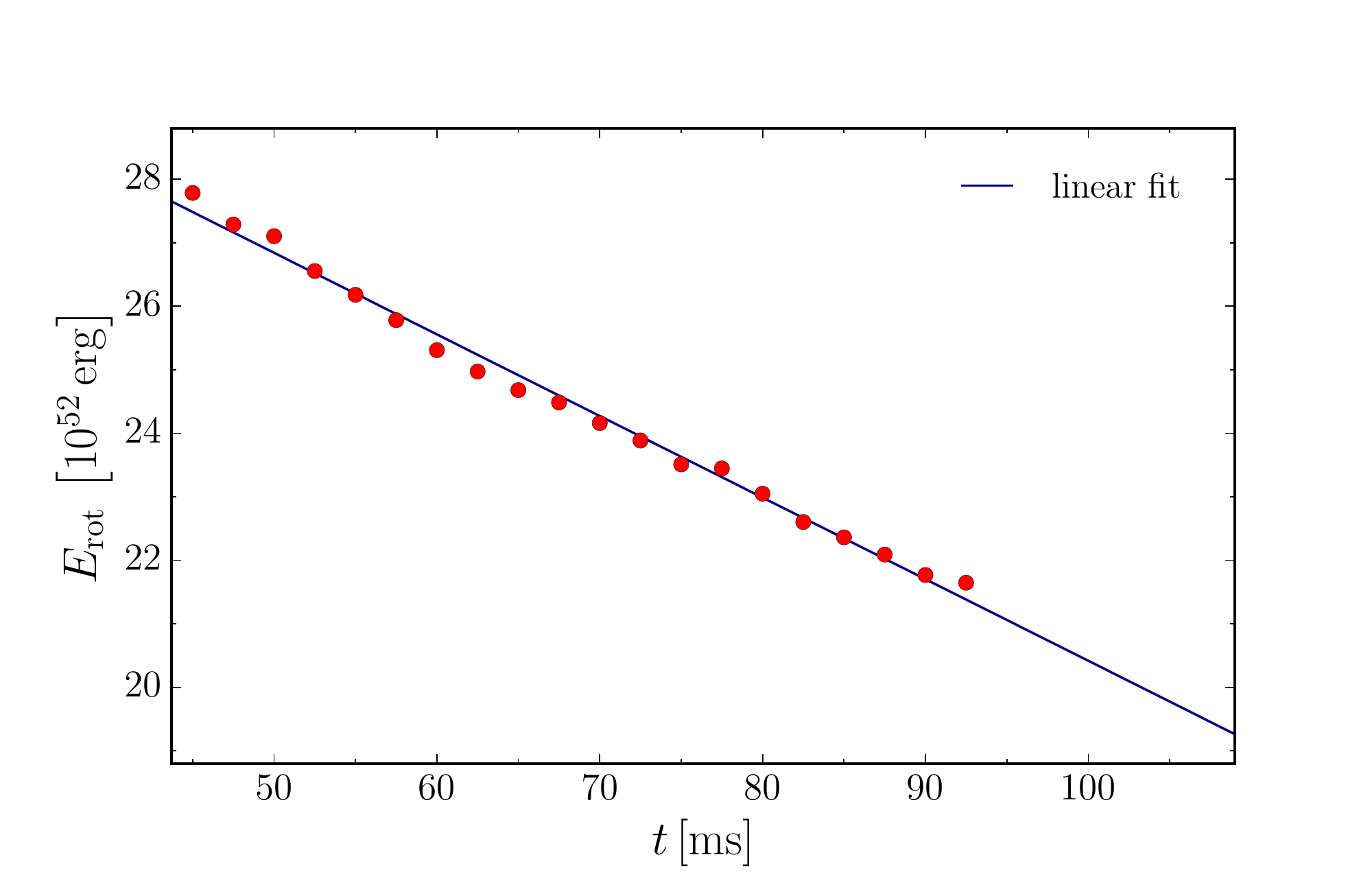}
  \caption{Rotational energy evolution (dots) along with a linear fit (continuous line).
  Merger time is $t=0$.} 
  \label{fig:Erot}
\end{figure}

We now consider the energy budget of the system.
The main energy reservoir available on subsecond timescales is the portion of rotational energy associated with differential rotation (see below). 
We compute the total rotational energy according to \cite{Shibata2000,Corvino2010}
\begin{equation}
E_\mathrm{rot}=\frac{1}{2}\int d^3 x \,\alpha \sqrt{\gamma}\,\Omega \,T_\phi^{\,0} \,\, ,
\end{equation}
where $\alpha$, $\gamma$, $\Omega$, $T^{\mu\nu}$ are lapse, 3-metric determinant, angular velocity, and stress-energy tensor. 
We find that the region defined by $z<10$\,km and cylindrical radius $r_\mathrm{cyl}<20$\,km always contains $>\!99$\% of $E_\mathrm{rot}$.
Figure~\ref{fig:Erot} shows the evolution of $E_\mathrm{rot}$ over the last 50\,ms, 
at which point the system has become highly axisymmetric.
From 50 to 100\,ms after merger the rotational energy decreases by 
$\Delta E_\mathrm{rot} \simeq 6\times10^{52}\,$erg~$\simeq 0.03\,M_\odot$. 
As discussed in Sec.~\ref{sec:remnant}, this decrease is likely dominated by 
the consumption of differential rotation and therefore the associated timescale should be determined by viscous effects. As a note of caution, the timescale found in our simulation is sensitive to the resolution employed, which determines how the physical mechanism responsible for the effective viscosity (e.g.,~magnetic fields and/or turbulence) competes with the numerical viscosity. 
Without a detailed resolution study, the observed timescale is not
reliable and a comparison between different magnetic field strengths
indicates that it might in fact be dominated by finite resolution
effects (see Sec.~\ref{sec:remnant}).

We find that most of the above $\Delta E_\mathrm{rot}$ is spent to rearrange and heat up the remnant NS, while only a small fraction is transferred to the surrounding matter. In particular, the kinetic energy associated with the outgoing radial motion of the matter outflow (at $r>300$\,km) increases by less than $10^{51}\,$erg [conservative upper limit assuming a radial velocity of 0.1\,c; Fig.~\ref{fig:rho_vel}(b)].
Moreover, magnetic energy variations are also smaller than $10^{51}$\,erg 
(Fig.~\ref{fig:Emag-Bmax}). 
In the following evolution, a similar amount of rotational energy can still be released before the remnant reaches uniform rotation (Sec.~\ref{sec:remnant}). It seems likely that this additional energy will be distributed in a similar manner.

The rotational energy at 100\,ms after merger is still $\simeq2\times10^{53}$\,erg, which would be in principle more than sufficient to produce a SGRB jet. In particular, the estimated jet core energy of GRB\,170817A is $\approx 4.4\times10^{49}$\,erg \cite{Ghirlanda2019}, almost 4 orders of magnitude smaller. 
However, the key question is whether a jet formation mechanism with sufficient efficiency exists and can be activated at later times. 
In this respect, the conditions at the end of our simulation are not promising. There is no accretion torus around the central remnant (Sec.~\ref{sec:remnant}), nor a strongly magnetized and low density funnel along the axis.    
Instead, we observe a slow and roughly isotropic outflow of matter, without any significant variation over a timescale of $\sim\!50$\,ms. 
In conclusion, we cannot conclusively rule out that the NS remnant could launch a jet at later times energetically compatible with GRB\,170817A, but our current findings point in disfavor of such a scenario.\footnote{Analogous conclusions were obtained from the long-term ($\sim\!200$\,ms) evolution of an axisymmetric and differentially rotating supramassive NS initially endowed with an internal and external dipolar magnetic field \cite{Ruiz2018}.} 

We note that in our simulation we are neglecting neutrino radiation. Neutrino-antineutrino annihilation along the spin axis could contribute to power an incipient jet \cite{Eichler:1989:126}. Nevertheless, the corresponding energy deposition would be $\lesssim10^{49}$\,erg (e.g.,~\cite{Perego2017}), insufficient to power a SGRB like GRB\,170817A even assuming 100\% energy conversion into $E_\mathrm{jet,\,core}$ and in general much smaller than the rotational energy budget. 
At the same time, neutrino emission and reabsorption would enhance the mass ejection via postmerger baryon winds, but the corresponding contribution would be negligible (e.g.,~\cite{Dessart:2009:1681}).
Therefore, we do not expect that the inclusion of neutrino radiation could significantly alter our results. 

In our model, we consider only one of the possible combinations of progenitor NS masses and NS EOS that are compatible with the GW signal of GW170817.\footnote{Note that our total mass differs by $\sim\!1\%$ from the one inferred for GW170817.} 
Different mass ratios and EOS would certainly have a substantial impact on the evolution of the remnant and even restricting only to those models that lead to a long-lived NS remnant, a qualitatively different result on jet formation remains possible. 

Concerning the dependence on the level of magnetization, it is not easy to make predictions. In order to create more favorable conditions for jet formation, we imposed a very large magnetic field, which is expected to enhance the jet power.\footnote{For instance, for the Blandford-Znajek mechanism the jet power scales as $\propto\!B^2$ \cite{Thorne1986}.}  
Nevertheless, we found that postmerger mass ejection is also strongly enhanced for a higher magnetization, leading to a much heavier obstacle opposing the propagation of an incipient jet.
The complex combination of the two effects makes it difficult to anticipate the outcome for different magnetizations.

We now examine the simulation in which the remnant was induced to collapse at 72\,ms after merger.
In this case, enough matter remains outside the BH to form an accretion disk of $\sim\!0.1\,M_\odot$ (see also Sec.~\ref{sec:remnant}).
The resulting BH-disk system might be able to launch an incipient SGRB jet via the Blandford-Znajek mechanism 
\cite{Blandford1977}, with power \cite{Thorne1986} 
\begin{equation}
L_\mathrm{BZ}\sim 10^{52} \Big{(}\frac{\chi}{0.5}\Big{)}^2  \Big{(}\frac{M_\mathrm{BH}}{2.5\,M_\odot}\Big{)}^2 \Big{(}\frac{B_\mathrm{BH}}{10^{16}\,\mathrm{G}}\Big{)}^2~\mathrm{erg/s}\, ,
\nonumber
\end{equation}
where we are using as reference values the dimensionless BH spin $\chi$, mass $M_\mathrm{BH}$, and magnetic field strength at the poles $B_\mathrm{BH}$ found in our simulation. 

Such a jet would have to drill through a very massive ejecta layer ($\sim\!0.1~M_\odot$).
By analyzing the conditions along the spin axis from $z=200$\,km (above which the matter distribution is not influenced by the collapse; Fig.~\ref{fig:rhodiff}) up to the ejecta front at 72\,ms after merger, we find that the jet energy density would be at any point orders of magnitude smaller than the rest-mass energy density of the ejecta (assuming a jet half opening angle $\theta_\mathrm{j}\gtrsim5^{\circ}$). This implies a nonrelativistic jet head velocity (e.g.,~\cite{Bromberg2011}). Following \cite{Bromberg2011} [Eqs.~(3) and (4a)], we roughly estimate the time required for the jet to break out as 0.01, 0.03, and 0.14\,s for a constant $\theta_\mathrm{j}$ of 5$^{\circ}$, 10$^{\circ}$, and 20$^{\circ}$, respectively. For $\theta_\mathrm{j}\gtrsim30^{\circ}$, the jet head would not be fast enough to ever reach the ejecta front. 
In conclusion, for a collimation of $\theta_\mathrm{j}\lesssim20^{\circ}$ and an engine duration $\Delta t_\mathrm{engine}\gtrsim0.15$\,s we should expect a successful jet despite the heavy ejecta. 
This result is also in agreement with the more recent estimates in \cite{Duffell2018}, which account for the expanding nature of the surrounding medium and are supported by numerical simulations.

The main factors that may affect the above conclusion are the remnant lifetime and the level of magnetization. A shorter lifetime would imply a lower ejecta mass and thus more favorable conditions for a jet breakout. A lower magnetic field strength, on the other hand, would lower both the ejecta mass and the jet power, which might act both in favor or disfavor.

Our present results leave ample space for an explanation of GRB\,170817A based on a BH central engine, in particular for remnant lifetimes $\lesssim\!100$\,ms.
Longer lifetimes would lead to a larger amount of ejecta, likely making the production of a successful jet increasingly more challenging. However, setting a reliable upper limit would require additional simulations exploring different magnetizations and a longer postmerger evolution.


\section{Remnant structure and rotation profile}
\label{sec:remnant}

\begin{figure}
  \begin{center}
    \includegraphics[width=0.999\columnwidth]{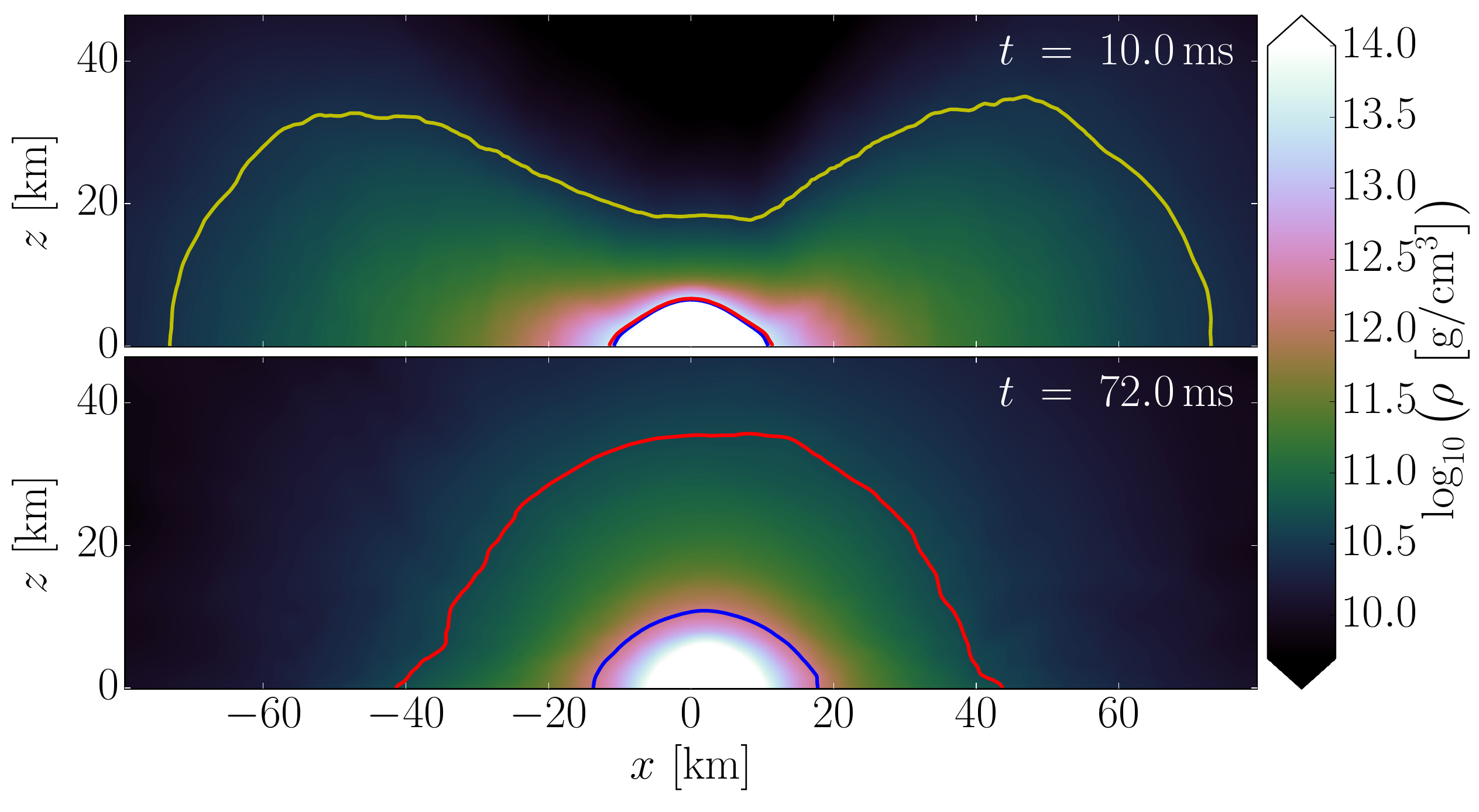}  
    \caption{Small scale meridional view of the rest-mass density at 10 and 72\,ms  
    after merger, in simulation coordinates. In order to remove contributions from 
    oscillations, we show the average over 4\,ms around the given times. 
    The contours correspond to isodensity surfaces that contain the fiducial 
    remnant baryon mass (blue; see text), 93\% of the total baryon mass (red), and 
    98\% of the total baryon mass (yellow, only visible in top panel).}
    \label{fig:remnant}
  \end{center}
\end{figure}

\noindent In this section, we focus the attention on the remnant properties at smaller scales. 
Figure~\ref{fig:remnant} shows the density distribution of the remnant 10 and 72\,ms after merger. 
At an early stage, an inner core is embedded in a torus-shaped outer envelope. 
This structure is typical for BNS merger remnants (e.g.,~Fig.~16 in \cite{Kastaun:2016}).
During the remainder of the simulation, we observe a general expansion. The matter in the torus occupies an increasing volume while assuming a more spherical configuration. This is shown in Fig.~\ref{fig:remnant}. 
 
As a quantitative measure for the radial mass distribution, we compute baryonic mass 
and proper volume enclosed in the surfaces of constant mass density, obtaining a gauge 
independent mass-inside-volume relation (see \cite{Kastaun:2017}). 
We will express proper volume by the radius of a Euclidian sphere of the same volume.

For the final state, we find a plateau in the mass-inside-radius relation that allows us to define a fiducial remnant baryon mass of $2.75\,M_\odot$. Most of the remaining baryon mass ($0.23\,M_\odot$) has reached radii between $100$ to $1000$ km at the end of the simulation.
It is worth noting that the outer layers of the central NS strongly expand.
The volume occupied by the fiducial remnant mass increases from an equivalent sphere radius of $14\,\mathrm{km}$ at $20\,\mathrm{ms}$ after merger to a radius of $31\,\mathrm{km}$ at $95\,\mathrm{ms}$. Toward the core, the change in radius becomes gradually smaller.

The expansion implies that the (negative) gravitational binding energy should increase; i.e.,~the NS becomes less strongly bound.
In this regard, we also refer to the discussion in \cite{Radice:2017:838} where it was demonstrated that prescribing a large turbulent viscosity can reduce the compactness of the remnant in the first few ms after merger via turbulent dissipation and angular momentum redistribution. 

\begin{figure}
  \begin{center}
    \includegraphics[width=0.999\columnwidth]{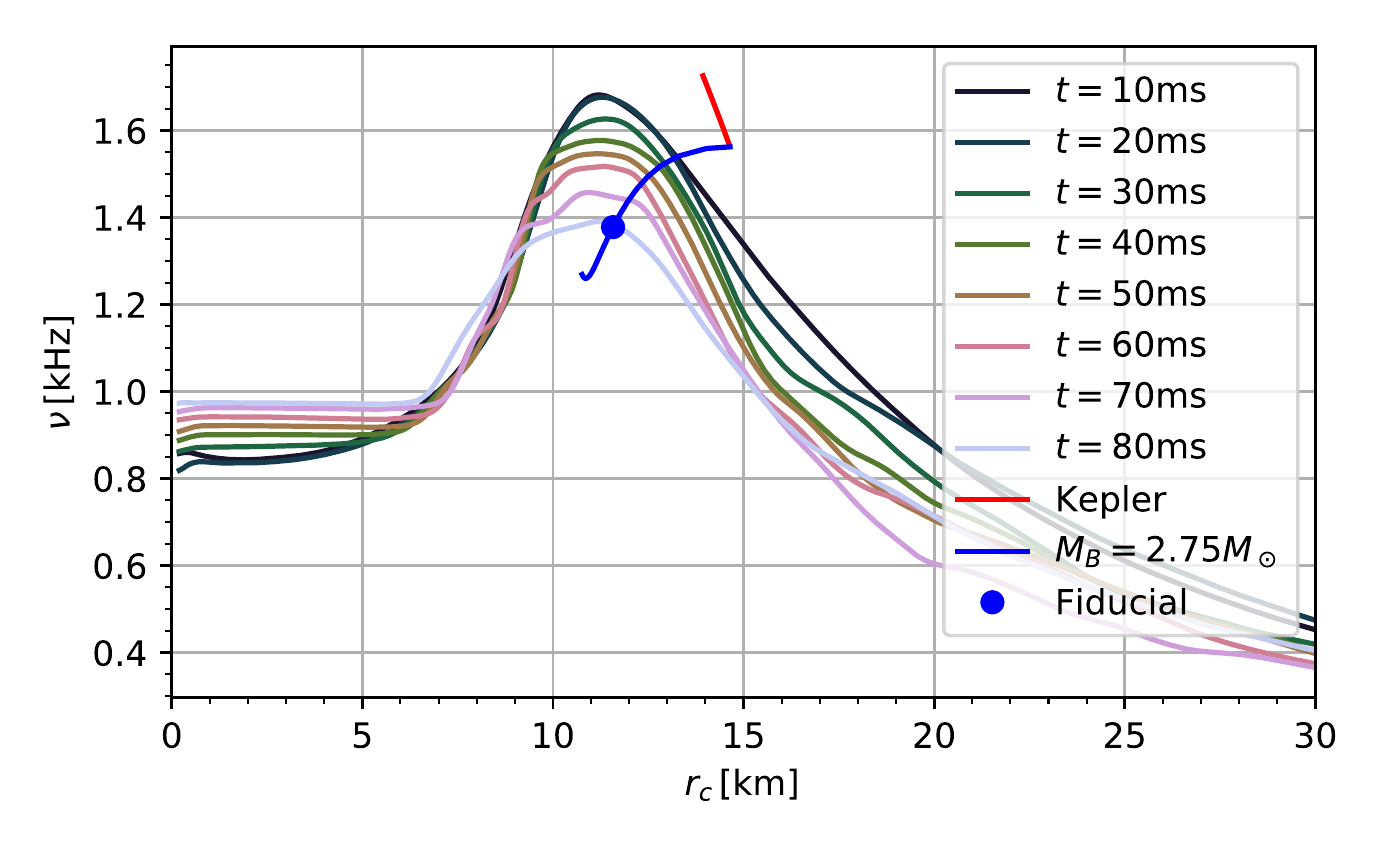}  
    \includegraphics[width=0.999\columnwidth]{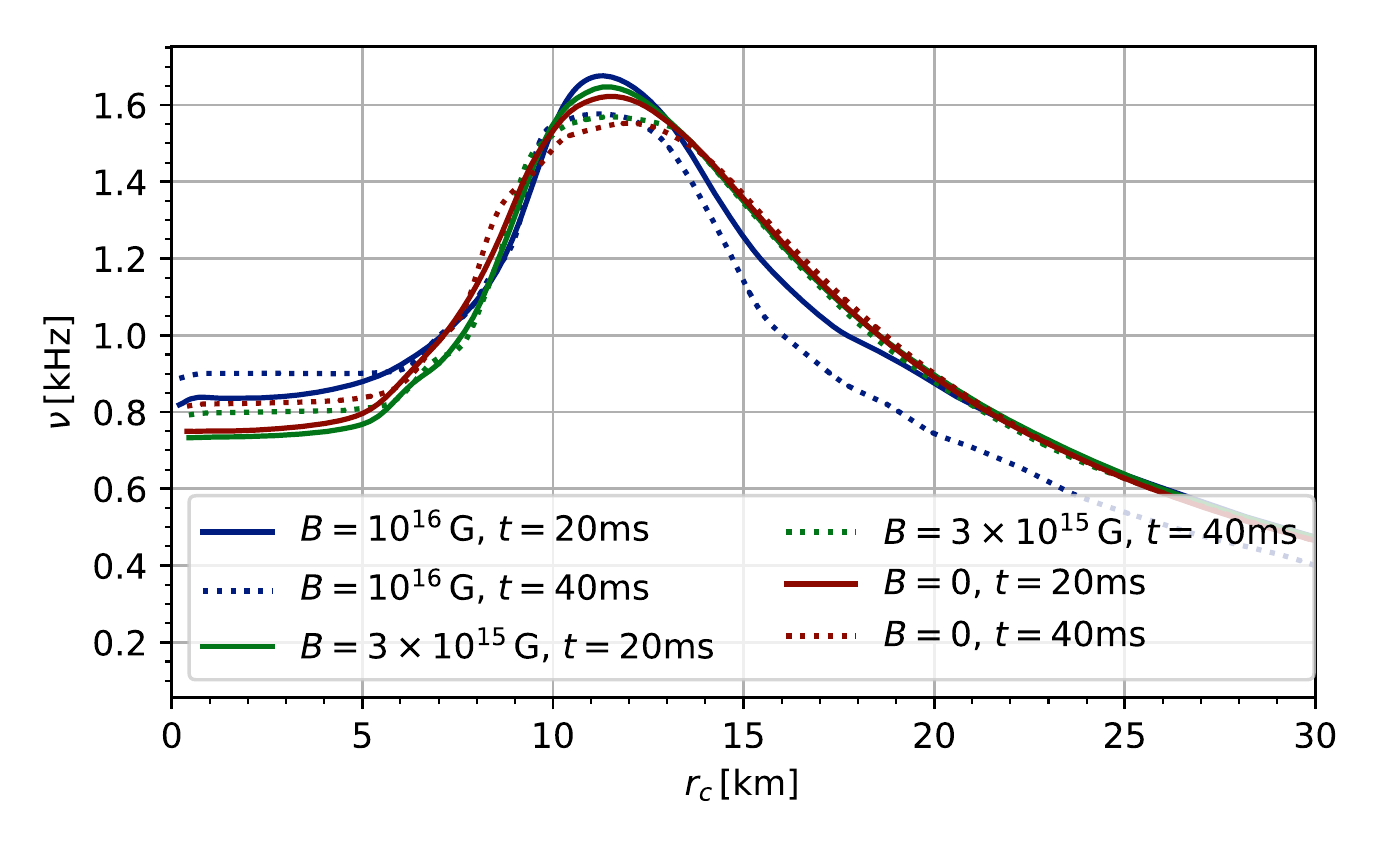}  
    \caption{{\it Top:} Evolution of the rotation profile in the equatorial plane. The curves represent the rotation rate averaged in the $\phi$ direction and over $2\,\mathrm{ms}$ in time 
    versus proper circumferential radius, at different times (values relative to merger time). For comparison, we show the rotation rate and circumferential radius of the uniformly 
    rotating fiducial model (see text), the uniformly rotating sequence of constant baryon mass (blue line), and the mass shedding sequence with baryonic masses between 
    fiducial remnant mass and total binary mass (red line). 
    {\it Bottom:} Comparison with the lower and zero magnetic field cases in \cite{Ciolfi2017} at 20 and 40\,ms after merger. The field strengths in the legend refer to the initial 
    maximum field strength of each model.}
    \label{fig:rot_prof_evol}
  \end{center}
\end{figure}

The rotational profile along the equator is shown in the upper panel of Fig.~\ref{fig:rot_prof_evol}.
As already noted in previous studies (e.g.,~\cite{Kastaun:2015:064027, Endrizzi:2016:164001, Kastaun:2016, Kastaun:2017, Hanauske2017, Ciolfi2017, Radice:2017:838}), the remnant is characterized by a slowly rotating central core (circumferential radius $r_\mathrm{c}\lesssim5$\,km) surrounded by a rapidly rotating outer layer. 
Further out, the centrifugal force starts dominating over the pressure and the rotation profile
approaches the Keplerian one. For the case at hand, the maximum spin frequency is around $1.7$ kHz, 
and the maximum is located at $r_\mathrm{c}\approx12$\,km.

For comparison, we computed the properties of uniformly rotating NS obeying the same zero-temperature
EOS as the initial data. Figure~\ref{fig:rot_prof_evol} shows rotation rate and radius for a model
with the baryonic mass of the fiducial remnant and its angular momentum at the end of the simulation. 
As one can see, the final rotation rate for uniform rotation is similar to the maximum rotation rate 80\,ms after merger. Comparing to the sequence of constant baryonic mass shown in the same figure,  
we find that the central rotation rate is much lower than the minimum allowed
uniform rotation rate for the given mass. At the same time, the remnant extends to larger
radii than possible for a uniformly rotating star.

The amount of differential rotation decreases during our simulation,
as shown in the upper panel of Fig.~\ref{fig:rot_prof_evol}, 
while the central core gains angular momentum at the expense of the outer envelope. 
We also find that a growing part of the core shows a flat rotation profile, i.e.,~uniform rotation.
The difference between maximum and central rotation frequencies, $\Delta\nu=\nu_\mathrm{max}-\nu_\mathrm{core}$, 
decreases from a maximum of $849 \,\mathrm{Hz}$ at $t=19 \,\mathrm{ms}$ after merger 
down to $340 \,\mathrm{Hz}$ at $t=91 \,\mathrm{ms}$.

There are different possible causes for the decrease. One is numerical dissipation which acts as 
a viscosity.
Physical causes include effective viscosity caused by strong small scale magnetic fields or turbulent motion,
as well as magnetic winding (see also \cite{Radice:2017:838,Shibata:2017:083005,Duez2006b}). 
We can assess the contribution of magnetic effects by comparing
simulations of the same system with different magnetizations. This is depicted in the lower panel of
Fig.~\ref{fig:rot_prof_evol}, which shows the evolution between $20$ and $40 \,\mathrm{ms}$ after 
merger.
We find that the magnetization has a mild influence on the initial rotation profile, and the 
decrease in $\Delta\nu$  is comparable for the three cases. 
This indicates that the differential rotation in the inner region is mostly dissipated
by nonmagnetic effects in our simulations. However, it does not imply that such effects are small 
in reality. In our setup, the small scale magnetic fields are most likely not sufficiently resolved
to account for the corresponding effective viscosity. We refer to \cite{Kiuchi:2018} for a detailed 
discussion based on simulations with much higher resolutions.

We note that in the outer envelope the magnetic field has a more substantial impact. 
The simulations with weaker and zero magnetic fields show almost no evolution in that region, 
while the present simulation reveals up to $\sim\!20$\% decrease in rotation frequency. 
Our simulation suggests that above $E_\mathrm{mag} \sim 10^{51}$\,erg, magnetic fields 
start to affect the outer layers within a timescale below $100$ ms.

Together with the differential rotation, the rotational energy shown in Fig.~\ref{fig:Erot} 
is decreasing as well. 
As discussed in Sec.~\ref{sec:sgrb}, the kinetic and magnetic energy in the surrounding 
matter increases only by a small fraction of the rotational energy loss. 
This indicates that rotational energy is converted into potential energy 
and heat.
The aforementioned expansion of the remnant supports this conclusion.

In order to obtain an estimate for the further rotational energy loss until 
uniform rotation is reached, we compute the rotational energy of the uniformly rotating
model with same baryonic mass and final angular momentum as the fiducial remnant. 
We find that a rotational energy around $8 \times 10^{52}$ erg would need to be 
converted (assuming no angular momentum loss) before reaching a uniformly rotating 
zero-temperature state. 
For comparison, the gravitational binding energy of the uniformly rotating 
model is $1.4 \times 10^{54}\,\mathrm{erg}$.

The uniformly rotating state can release further energy if we allow for angular momentum loss,
e.g.,~during magnetic spin-down. It will collapse to a BH when the minimum angular momentum for the 
given mass is reached, at a rotation rate around $1.3$ kHz.
The total energy (gravitational mass) of the uniformly rotating NS would decrease by $3\times 10^{52}$ erg 
during the spin-down. This energy reservoir can only be used on long timescales, however, since it requires 
shedding a large amount of angular momentum ($\sim\!0.4\,GM_\odot^2/c$).

So far, we discussed the long-term development of the remnant without considering possible fallback of 
the outgoing matter. Most of it remains bound according to the geodesic criterion, which, however, does not 
account for the large degree of magnetization. It is therefore instructive to discuss a hypothetical
scenario where matter falls back, circularizes by self-interaction, and forms an accretion disk.
An important aspect for the central remnant evolution is the amount of angular momentum and mass
added by accretion. 
Accretion is driven by an outward flow of angular momentum caused by effective 
viscosity. The divergence of the flow reduces the specific angular momentum of fluid elements, causing 
them to migrate inwards. The total angular momentum at the end of our simulation 
is slightly larger than the maximum possible for a uniformly rotating NS with the same total baryonic  
mass. However, given the angular momentum transport during accretion, we do not expect the rotation rate 
of the central remnant to approach the mass shedding limit (note that this is contrary to the
assumptions made in another work on the final state of merger remnants \cite{Radice:2018:481}).
The rotation rate and radius of the mass shedding sequence up to the total mass of the system are shown in 
Fig.~\ref{fig:rot_prof_evol} for comparison.
\begin{figure}[!t]
  \begin{center}
    \includegraphics[width=0.95\columnwidth]{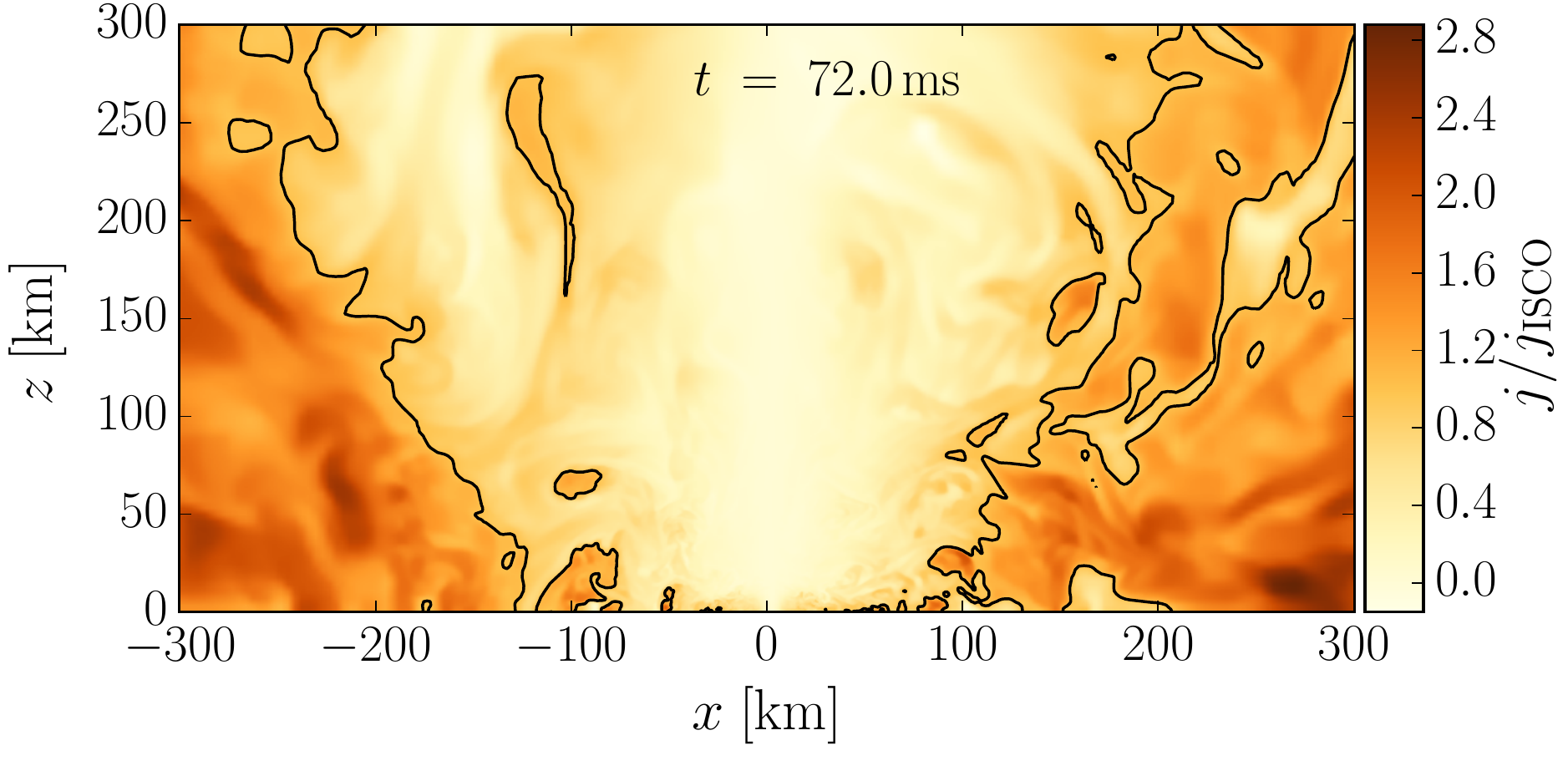}  
    \caption{Meridional view of the specific angular momentum with respect to the 
    oribital (or remnant spin) axis, 72\,ms after merger. The value is given in units of 
    the specific angular momentum of a test particle on the innermost stable circular 
    orbit around the black hole formed in the collapsing case
    ($j_\mathrm{ISCO}$). The black contour corresponds to $j=j_\mathrm{ISCO}$.}
    \label{fig:jISCO}
  \end{center}
\end{figure}   

To get an upper limit for the angular momentum of the central remnant after
accreting a certain mass, we ignore the outwards angular momentum flow between mass shells and assume
that the accreted mass contributes the specific angular momentum corresponding to some radius close to the 
remnant. Since the central NS is in the supramassive mass range, stability requires a minimum angular
momentum that increases when adding mass. Starting from the fiducial values for mass and angular momentum,
we can compute the angular momentum that needs to be contributed by the accreted matter
to prevent collapse. 

We find that the required average specific angular momentum raises from zero for a total accreted 
mass of $M_\mathrm{acc}\approx0.05\,M_\odot$ to $J/M_\mathrm{acc}\approx 5.8\,GM_\odot/c$ when 
accreting all of the outflowing matter. The specific angular momentum in orbit around the equator of
the minimally rotating supramassive NS sequence is below $4.6\,GM_\odot/c$. It is therefore possible
that the remnant collapses due to an insufficient increase of angular momentum before reaching the 
maximum mass of a rotating NS, leaving behind an accretion disk of up to $\sim\!0.2\,M_\odot$.

We now consider the case in which we induced collapse to a BH. We are interested, in particular, in the mass of the accretion torus left around it.
According to a widely used criterion, only matter fulfilling $j>j_\mathrm{ISCO}$ before the collapse could 
contribute to the accretion torus, where $j_\mathrm{ISCO}$ is the specific angular momentum of a test 
particle on the innermost stable circular orbit around the newly formed BH.  
Guided by such a criterion, recent studies addressed the problem by considering idealized models of uniformly rotating supramassive NSs and their collapse to a BH \cite{Margalit2015,Camelio2018}.
Their findings suggest that even assuming maximally rotating NS configurations, the resulting torus mass would be most likely $\ll\!0.1\,M_\odot$ and thus insufficient for a BH-powered SGRB jet. 

For the BH formed in our simulation,  $j_\mathrm{ISCO}c/GM\simeq 2.9$ (computed following \cite{Bardeen1972}). 
Right before the collapse only a negligible mass fraction satisfies $j>j_\mathrm{ISCO}$ within a radius of $\simeq\!50$\,km (Fig.~\ref{fig:jISCO}). 
A simple application of the above criterion would then imply that at least all the matter within $50$\,km from the remnant center should be directly swallowed by the BH. However, only $\lesssim\!97$\% of the corresponding baryon mass is actually swallowed. 
Such a discrepancy can be explained by the fact that part of the material surrounding the remnant has a significant outgoing velocity, deviating from the conditions for which the criterion is supposed to hold. Although $\approx3\%$ might seem a very small difference, this is sufficient to have a massive ($\sim0.1\,M_\odot$) accretion torus and thus a potentially viable SGRB central engine.

This example shows that the bulk of the NS remnant satisfies the condition $j<j_\mathrm{ISCO}$ much before achieving uniform rotation. Nevertheless, this condition by itself does not prevent the formation of a massive accretion torus in case of a collapse.
The conclusion that the resulting BH could not act as a SGRB central engine only applies when the remnant NS is no longer surrounded by baryon wind material that could contribute to form an accretion torus.
Given that in our simulation no significant change in density distribution is observed up to 400\,km in the last 50\,ms (Fig.~\ref{fig:rhofunnel}), the timescale required to reach the latter condition should be $\gg\!100$\,ms.


\section{Gravitational waves and remnant oscillations}
\label{sec:oscil}

\noindent In the following we discuss the remnant oscillations and the resulting GW signal.
Thanks to the long evolution time, we can study the late time behavior in addition 
to the usual inspiral and early postmerger phase. 
We will also discuss how the GW signal relates to the evolution of the remnant.

\begin{figure}
  \begin{center}
    \includegraphics[width=0.95\columnwidth]{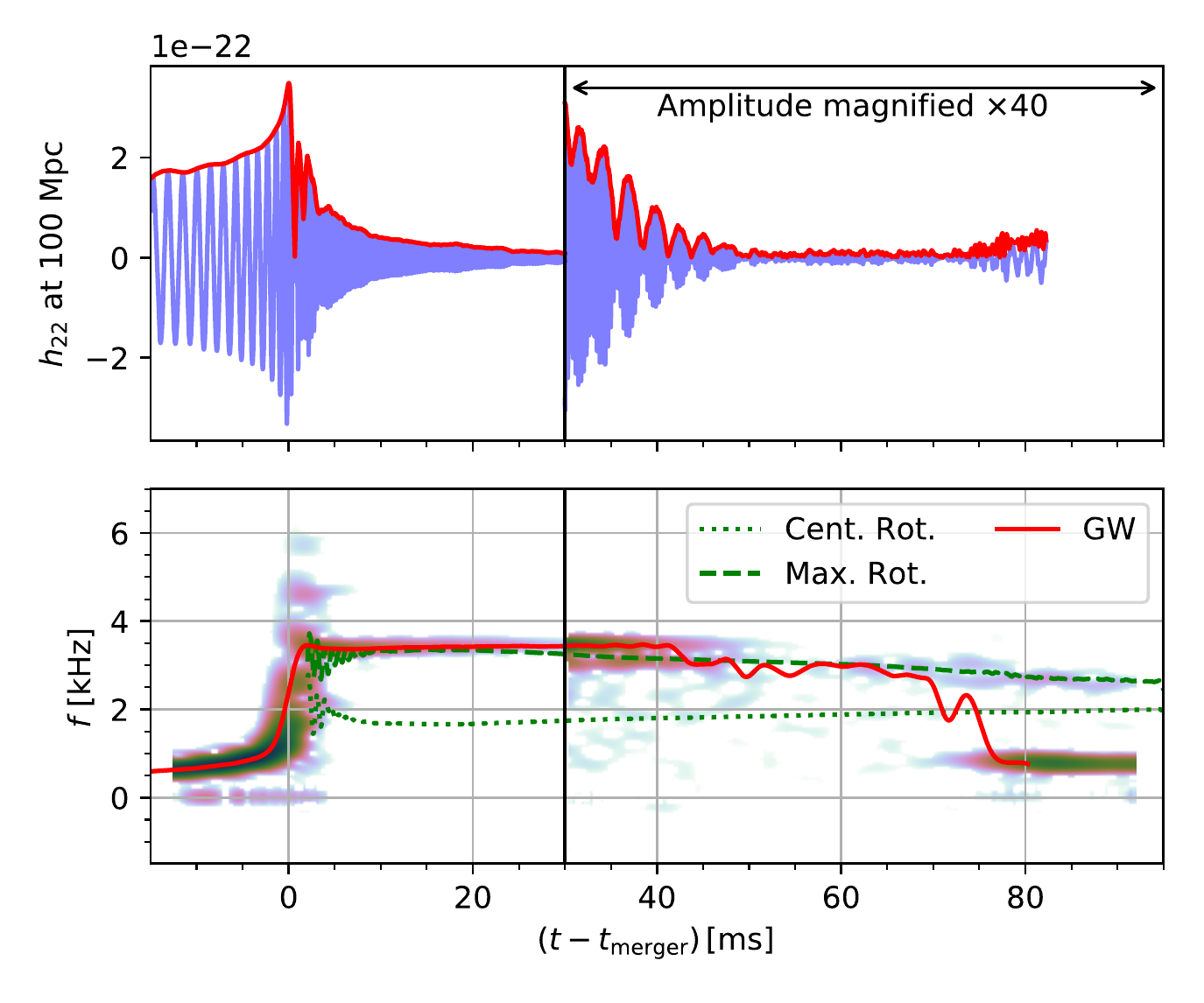}  
    \caption{{\it Top}: GW strain $l=m=2$ multipole coefficient (real part).
    The strain after $t=30$ ms is magnified by a factor 40.
    {\it Bottom}: The red curve shows the phase velocity of the GW strain. 
    The green curves show twice the central (dotted line) and twice the maximum 
    rotation rates (dashed line).
    The color plot shows a spectrogram of the $m=2$ component of 
    the density perturbation in the equatorial plane averaged in the whole 
    remnant. The scale is logarithmic, spans 5 orders of magnitude,
    and the amplitude after $t=30$ ms is scaled by a factor of 1000.
    All phase velocities are smooth averages over $1$ ms. 
    Times are with respect to merger time (and using retarded time for 
    the GW signal). Note the oscillation at the end of the simulation
    is likely an artifact.
  }
    \label{fig:gwspecgram}
  \end{center}
\end{figure}

Figure~\ref{fig:gwspecgram} shows the evolution of the GW strain,
which is fairly typical for a BNS system in the supramassive mass range.
The amplitude is decaying strongly after merger. The total radiated energy
during the evolution from $10\,\mathrm{ms}$ after merger onward is only
$4.6 \times 10^{-3} \, M_\odot$, which is ${\sim}0.5$\% of the binding energy
of the fiducial remnant (see Sec.~\ref{sec:remnant}). 
Less than 2\% of the total angular momentum is carried away by GW during this period.
The GW emission therefore has only a minor impact on the evolution of the remnant.

For the early postmerger phase, up to around $30$ ms after the merger, we confirm the 
results of earlier numerical studies \cite{Kastaun:2016,Kastaun:2017,Ciolfi2017}, which 
observe a gravitational wave consisting of a single component with a frequency equal to 
twice the maximum rotation rate. For our simulation, this is shown 
in Fig.~\ref{fig:gwspecgram}, comparing the evolution of the GW frequency to
the evolution of the maximum rotation rate in the equatorial plane. 
During this period, the GW amplitude decays strongly. 

In the subsequent evolution,
the qualitative behavior seems to change, as we will now discuss.
We caution that the numerical accuracy of oscillations with very small amplitudes
is not well studied, and we report the results as seen.
In the numerical results, we observe a splitting of modes around $30$ to $50$ ms 
after merger. Figure~\ref{fig:gwspecgram} shows a spectrogram
of the average $m=2$ component of the density in the equatorial plane,
extracted using the formalism presented in \cite{Kastaun:2015:064027}.
One can see a superposition of a mode with frequency following
twice the rotation rate and a component remaining at a slightly larger frequency.
Consequently, the GW strain exhibits a beating phenomenon visible in
the upper panel of Fig.~\ref{fig:gwspecgram}.

The cause of the split and the nature of the oscillations is currently unknown.
It might be related either to the decreasing differential rotation, which is
shown in Fig.~\ref{fig:gwspecgram}, or to the decaying amplitude of the perturbation.
It is possible that the nonaxisymmetric deformation in the early postmerger phase
is a genuinely nonlinear quasistationary state, as proposed in 
\cite{Kastaun:2016,Kastaun:2017}. With decreasing amplitude, the evolution of 
the remnant deformation might be governed more by linear perturbation theory. 
It is worth noting that, for a different model, the authors of \cite{DePietri2018} also observed the appearance of a 
second oscillation mode with frequency slightly lower than the mode which is dominant 
right after merger, and linked its excitation to a convective instability.

Near the end of the simulation, starting at around $70$ ms after merger, the extracted
GW signal exhibits a growing low-frequency oscillation. This is likely a numerical 
artifact, since we observe a growing oscillation
of the coordinate system as well, at the same frequency. Since the strain amplitude is very 
small at this time, it might well be an error induced by the behavior of the gauge.


\section{Summary and Conclusions}
\label{sec:conclusion}

\noindent We simulated a magnetized BNS merger forming a long-lived NS remnant and followed its evolution up to $\approx\!100$\,ms after merger.
This represents the longest postmerger evolution performed in GRMHD for such a system, allowing us to study the remnant properties over timescales
so far unexplored.

Starting from a rather large initial magnetic energy ($\sim\!10^{48}$\,erg), we observe further magnetic field amplification over the different merger phases up to a saturation level of $E_\mathrm{mag}\simeq 2\times 10^{51}$\,erg, reached around 50\,ms after merger. 
In the following $50$\,ms, magnetic energy shows no significant change, despite the fact that the dominant amplification mechanism (i.e.,~the MRI) is well resolved. This is indicative of a physical saturation. 

Within 90 ms of postmerger evolution, a large scale magnetic field of mixed poloidal-toroidal structure 
emerges, with magnetic strength reaching $\gtrsim\!10^{15}$\,G.
Since we did not impose initial magnetic fields extending outside the two NSs, the production of such a large scale field is entirely due to the postmerger dynamics. 
Our result supports the idea that BNS mergers can produce remnant NSs endowed with a global field of magnetar-like field strength.

The strongest field lines wind around the axis roughly along a wide cone and reach distances up to
${\approx}100$\,km from the equatorial plane. At the end of the simulation, those strong field lines 
still loop back, which makes them unsuitable for guiding an incipient jet.  
Furthermore, during the entire postmerger evolution, the remnant NS is surrounded by a dense environment 
with a nearly isotropic and steady rest-mass density distribution up to $\approx400$\,km.
Under those conditions, it is not surprising that we find no signs of jet formation during our simulation.

The strong magnetic field drives a slow ($<0.1\,c$) and massive postmerger outflow corresponding to a mass loss rate of up to $\dot{M}\!\sim3\,M_{\odot}/$s, for a total mass of 0.1\,$M_{\odot}$ at the end of the simulation. A significant fraction of this material might become unbound at larger radii, potentially leading to a very massive ejecta component.
Comparing with a lower magnetic field simulation, we find that this magnetically driven outflow has a strong dependence on the system magnetization: $M_{\odot}\propto(E_\mathrm{mag})^{\,a}$ with $1\!\lesssim\!a\!\lesssim2$. 
At large scales, we can clearly distinguish a faster outflow component aligned with the remnant spin axis (of half opening angle $\sim\!40^{\circ}$). This component is highly magnetized (typical magnetic-to-fluid pressure ratio between 0.1 and 10), but also too heavy to be accelerated up to relativistic speeds and turn into a jet. 
Instead, it represents an obstacle for any incipient jet possibly launched by the remnant at later times. 

Studying the energetics of the system, we find that the largest power source on subsecond timescales after merger
is the decrease of rotational energy by around $6\times 10^{52}$ erg from 50 to 100 ms after merger.
The major part of this energy corresponds to the decrease of differential rotation.
In the same time period, GWs carry away only around ${\sim} 8 \times 10^{51}$\,erg and 2\% of 
the angular momentum. 
Besides the small GW contribution, almost all of the rotational energy decrease 
is converted internally within the remnant, which expands (due to rearrangement of angular 
momentum and heating) and therefore becomes less strongly bound. 
In comparison, both the variation of total magnetic energy ($\lesssim\!10^{51}$ erg) and 
the increase of kinetic energy associated with the radial motion of the outflowing 
material ($<\!10^{51}$ erg) are negligible.

It is worth pointing out that the above kinetic and magnetic energy of the surrounding  matter 
are still large compared to the inferred energy of a SGRB like GRB\,170817A. Nevertheless, 
on the timescale of our simulation, there is no indication for a sufficiently efficient conversion 
mechanism and there is no accretion torus around the remnant NS, which excludes an accretion-powered jet.
On the contrary, we observe a slow, noncollimated, and rather steady matter outflow. 
Since we could not prove that all this matter will eventually become unbound, accretion
of fallback material on timescales $\gg\!100$ ms remains, however, a possibility.
In conclusion, our current results seem to disfavor, but cannot rule out, 
a scenario in which a long-lived NS remnant acts as a SGRB central engine. 

Based on previous studies (e.g.,~\cite{Perego2017,Dessart:2009:1681}), we do not expect that the above conclusion could be altered by introducing neutrino radiation. On the other hand, employing a different mass ratio or EOS within the range compatible with the GW signal of GW170817 could have a significant impact. 

The rotational profile on the equatorial plane shows a slower and uniformly rotating core and a faster rotating outer layer with the maximum spin frequency lying at a cylindrical radius of $\sim\!12$\,km. At larger distances, the angular velocity is close to a Keplerian profile, indicating that this part of the remnant NS is centrifugally supported. 
We also find that the GW frequency is twice the maximum rotation rate, as in earlier studies.
As the system evolves, the maximum spin frequency decreases and the central one increases. 
We estimate that the remnant present at the end of our simulation could reach uniform rotation  
after a further rotational energy loss of $\approx8\times10^{52}$\,erg, 
which is comparable to the amount lost over the last 50\,ms of the simulation. 

By comparing the rotational profile evolution at different magnetizations, we find that magnetic effects become important 
in the matter surrounding the remnant (over a 100\,ms timescale) for $E_\mathrm{mag}\gtrsim10^{51}$\,erg. 
However, the inner region of the remnant appears to be marginally affected even at the highest magnetizations. This indicates that in our simulation the effective viscosity driving the removal of differential rotation in the inner core is of numerical origin (i.e.,~dominated by finite resolution effects rather than magnetic fields or turbulence).
We emphasize that current numerical simulations cannot reliably determine the timescale for the dissipation of differential rotation in the remnant since the computational
costs for a simulation as long as ours with the required resolution would become extremely prohibitive.

In parallel to our long-lived NS simulation, we also performed a second simulation where we induced the collapse of the remnant $\sim\!72$\,ms after merger by altering the EOS at supranuclear densities.
The collapse leads to the formation of a $\sim\!2.5\,M_{\odot}$ BH with $Jc/GM^2\simeq0.5$ and surrounded by a $\sim\!0.1\,M_{\odot}$ accretion torus.
Only matter within $\sim\!200$\,km distance is influenced by the collapse, with the rest of the system essentially unaffected. 

Prior to collapse, all the matter within 50\,km from the core of the remnant NS has specific angular momentum lower than the value at the innermost stable circular orbit around the newly formed BH. 
Nevertheless, a massive ($\sim\!0.1\,M_{\odot}$) accretion torus is still produced. 
This can be explained by the fact that the NS remnant is surrounded by a dense environment, characterized by a non-negligible outward directed radial velocity. 
This example shows that the maximum collapse time after which no massive accretion torus can be produced depends on the timescale over which the remnant NS is no longer surrounded by such a dense environment. This timescale is most likely $\gg\!100$\,ms and could be as long as a few seconds or more. 

At collapse, about 80\% of the total magnetic energy is swallowed by the BH. 
Nevertheless, the magnetic field strength close to its poles remains very large ($\sim\!10^{16}$\,G). Thanks to the presence of a massive accretion torus, the conditions to launch a powerful SGRB jet via the Blandford-Znajek mechanism are in principle fulfilled. 
Moreover, we find that for a jet half opening angle of $\lesssim\!20^{\circ}$ and an engine duration of $\gtrsim0.15$\,s such a jet should be able to successfully break out of the thick ejecta layer, despite the large mass of the latter ($\sim\!0.1\,M_\odot$). This supports the possibility of a BH-powered SGRB jet compatible with the inferred properties of GRB 170817A.
Different remnant lifetimes and magnetization levels could, however, alter this conclusion. 

The most important results obtained in the present work were made possible by following about $100$\,ms of evolution of the long-lived NS remnant, which is more than a factor of 2 longer than previous GRMHD simulations of this kind \cite{Ciolfi2017}. This demonstrates how a significant extension of the physical time of current simulations can lead by itself to novel insights.  
On the other hand, our study is limited to a single BNS model. A long-lived NS remnant remains a likely outcome for a potentially large fraction of all BNS mergers, and thus in preparation for future detections we urgently need more simulations exploring the relevant parameter space (different NS masses, EOS, magnetization, etc.).


\section*{ACKNOWLEDGMENTS}

\noindent 
J.\,V.\,K.~kindly acknowledges the Ca.Ri.Pa.Ro Foundation (https://www.fondazionecariparo.it) for funding his Ph.D.~fellowship within the Ph.D.~School in Physics at the University of Padova. 
The numerical simulations were performed on the cluster MARCONI at CINECA (Bologna, Italy). We acknowledge CINECA awards under the ISCRA and MoU INAF-CINECA initiatives, for the availability of high performance computing resources and support (Grants IsB15\_ShortGRB and INA17\_C2A11).
In addition, part of the numerical calculations have been made possible through a CINECA-INFN agreement, providing access to further resources on MARCONI at CINECA (allocation INF18\_teongrav).


%


\end{document}